\documentstyle{article}
\include{epsf}

\begin{document}

\title{Time correlations and 1/f behavior in backscattering radar reflectivity
measurements from cirrus cloud ice fluctuations}

\author{ K. Ivanova,$^1$ T.P. Ackerman,$^2$ E.E. Clothiaux,$^1$ \\ P.Ch.
Ivanov,$^3$ H.E. Stanley,$^3$ \\and \\ M. Ausloos$^4$\\
$^1$ 
Department
of Meteorology\\ Pennsylvania State University\\ University Park, PA 16802, USA\\
$^{2}$ Pacific Northwest National Laboratory\\ U.S. Department of Energy,
Richland, WA 99352, USA\\ $^3$ Center for Polymer Studies, Boston University\\
Boston, MA 02215, USA \\ $^4$ SUPRAS and GRASP, University of Li\`ege\\ Sart
Tilman B5, B-4000 Li\`ege, Euroland}

\maketitle

\begin{abstract} The state of the atmosphere is governed by the 
classical laws of
fluid motion and exhibits correlations in various spatial and temporal scales.
These correlations are crucial to understand the short and long term trends in
climate. Cirrus clouds are important ingredients of the atmospheric boundary
layer. To improve future parameterization of cirrus clouds in climate 
models, it
is important to understand the cloud properties and how they change within the
cloud. We study correlations in the fluctuations of radar signals obtained at
isodepths of $winter$ and $fall$ cirrus clouds. In particular we focus on three
quantities: (i) the backscattering cross-section, (ii) the Doppler velocity and
(iii) the Doppler spectral width. They correspond to the physical coefficients
used in Navier Stokes equations to describe flows, i.e. bulk modulus, 
viscosity,
and thermal conductivity. In all cases we find that power-law time correlations
exist with a crossover between regimes at about 3~to 5~min. We also find that
different type of correlations, including $1/f$ behavior, characterize the top
and the bottom layers and the bulk of the clouds. The underlying mechanisms for
such correlations are suggested to originate in ice nucleation and 
crystal growth
processes. \end{abstract}


\section{Introduction}

Ice research spans a broad range of subjects, including studies of extremely
complex phase diagrams [{\it Mishima and Stanley}, 1998], and phase transition
modelling [{\it Huckaby et al.} 2000], hot topics as still seen recently [{\it
Putrino and Parrinello}, 2002] in $ab$ $initio$ calculations, or in various
properties, such as the dielectric constant [{\it Gingle and Knasel}, 1975] or
acoustic emission features [{\it Weiss and Grasso}, 1997] during crack
propagation. In clouds, ice crystals appear in a variety of forms and shapes,
depending [{\it Rogers}, 1976; {\it Heymsfield and Platt}, 1984] on 
the formation
mechanism and the atmospheric conditions. In cirrus clouds, at 
temperatures lower
than about $-40^{\circ}$~C ice crystals form and exist as mainly nonspherical
particles [{\it Petrenko and Whitworth}, 1999].

In air transport, ice covering of airplane wings is also known to be very
annoying. Moreover high altitude clouds, found above ca. 4~km including
stratosphere cirrus [{\it Tabazadeh et al.}, 1997] at $>20$~km, play 
an important
role in regulating the energy budget [{\it Wallace and Hobbs}, 1977] 
of the earth
atmosphere system through their interaction with solar and 
terrestrial radiation
[{\it Stephens}, 1990]. Their impact on the climate and weather systems at
various time and space scales is not well understood. Recent studies suggests
that the water vapor itself rather than the {\it greenhouse gases} could be the
culprit in the global warming [{\it Maurellis}, 2001; {\it Rosenfeld and
Woodley}, 2001]. Indeed due to the presence of water everywhere and the ability
to change its state from, say, its frozen form at the poles to its liquid and
vapor states in the atmosphere, the water is an effective transfer 
energy medium
around the globe. No need to say that the description of ice crystal motions in
clouds, and the cloud motion itself imply solving Navier-Stokes equations
containing physical coefficients not well determined at this time.

The radiative properties of the ice particles are important ingredients of the
parameterization algorithms [{\it Heymsfield and Platt}, 1984; {\it 
Sundqvist et
al.}, 1989; {\it Zhang and McFarlane}, 1995; {\it McFarquhar and Heymsfield},
1997; {\it Zurovac-Jevtic}, 1999] used in radiative transfer schemes of cirrus
clouds in large-scale climate models. The question is still open on how cirrus
radiative transfer properties might be predicted given their (actually) roughly
known microphysical characterization [{\it Smith and Del Genio}, 2001].
Therefore, it is essential that correlations between the structure of cirrus
clouds and the radiative properties be better understood. An 
interesting study on
composite cirrus morphology and correlations with temperature and large-scale
vertical motion has been also conducted in e.g. [{\it Mace}, 1997].

Because of the vertical extent, ca. from about 4 to 14~km and higher, and the
layered structure of such clouds, one way of obtaining some information about
their properties is using ground-based remote sensing instruments. (Satellites
can also be used [{\it Liou et al.}, 2002]).  Experienced workers at airport
radar sites have some $feeling$ about ice cloud content. However a scientific
inner cloud morphology investigation consists in searching for the statistical
properties of radio wave signals backscattered from the ice crystals present in
the cloud. Backscattered signals received at the radar receiver 
antenna are known
to depend on the ice mass content and on the particle size distribution [{\it
Atlas et al.}, 1995]. Ice particles can induce modifications in the 
radar signals
which lead to fluctuations in (i) the backscattering cross section, (ii) the
Doppler velocity and (iii) the Doppler spectral width. The motivation for our
study is to identify properties of the inner structure of the cirrus clouds
through a statistical analysis of the fluctuations of such measured 
signals. This
is a first step toward including these statistics in parameterization 
schemes of
cirrus clouds in large-scale climate models (see references quoted supra) or in
short scale Navier-Stokes equation solutions. It is of interest to examine the
$time$ correlations in the backscattered signals not only on the 
boundary layers,
i.e. at the top and bottom of the cloud, but also at several levels within the
cloud, because of the extended vertical structure of cirrus clouds. Analysis of
$spatial$ correlations in vertical direction will be considered in a future
study.

Due to the nonlinear physics laws governing the phenomena in the 
atmosphere, data
series of atmospheric quantities are usually $non-stationary$. Therefore a mere
study based on the power spectrum only would be incomplete, if not misleading.
New techniques have been recently developed that can systematically eliminate
trends in the data and thus reveal some intrinsic dynamical properties such as
fluctuation correlations that are very often masked by nonstationarities. There
have been indeed a large number of studies on long-range power-law correlations
in time series [{\it Kantz and Schreiber}, 1997; {\it Brockwell and 
Davis}, 1998;
{\it Malamud and Turcotte}, 1999; {\it Schreiber}, 1999] in many 
research fields
such as biology [{\it Stanley et al.}, 1993; {\it Hausdorff et al.}, 1997; {\it
Ivanov et al.}, 1996, 1999; {\it Mercik et al.}, 2000; {\it Ashkenazy et al.},
2001], finance [{\it Mantegna and Stanley}, 1995; {\it Vandewalle and Ausloos},
1997; {\it Ausloos and Ivanova}, 1999] meteorology and climatology [{\it
Koscielny-Bunde et al.}, 1993, 1998; {\it Davis et al.}, 1996; {\it Ivanova et
al.}, 2000, 2002; {\it Ausloos and Ivanova}, 2001; {\it Bunde et 
al.}, 2001; {\it
Peters et al.}, 2002 ]. Some of us have already examined long-range time
correlations in nonstationary atmospheric signals of the liquid water 
path [{\it
Ivanova et al.}, 2000] and water vapor path [{\it Ivanova et al.}, 2002] in
stratus clouds, as well as in large-scale meteorological signals like the
Southern Oscillation Index [{\it Ausloos and Ivanova}, 2001].

We pursue the analysis of millimeter wave cloud radar data [{\it Mace et al.},
1997, 1998], hereby collected from observations during two consecutive $winter$
days, i.e. Jan. 26 and 27, 1997 and one $fall$ day, i.e. Sept. 26, 1997 at the
Southern Great Plains site of the Atmospheric Radiation Measurements (ARM)
program. Mace et al., [1998] have derived the microphysical 
properties of cirrus
layer from surface based millimeter radar and infrared radiometer data. Mace et
al. [1998] have developed techniques for retrieving cirrus cloud 
boundaries, ice
water content, crystal effective radius, and number concentration from
reflectivities combined with atmospheric emitted radiance interferometer
downwelling radiances and have applied the algorithm to almost two 
years of MMCR
data (11/96-5/98) at the Atmospheric Radiation Measurement (ARM) 
Program Southern
Great Plains (SGP) Cloud and Radiation Testbed (CART) site. Here we analyze the
statistical properties of the three quantities: (i) the backscattering cross
section per unit volume, (ii) the Doppler velocity and (iii) the 
Doppler spectral
width, obtain their time correlations, and give some physical interpretation.

First, let ${\cal N}(D, \vec r)$ be the distribution function for the number of
particles per unit volume; each particle is supposed to have a characteristic
size $D$ when located at $\vec r$; let $\sigma(D)$ be the backscattering cross
section of particles with such a characteristic size $D$. Then the total
backscattering cross section (BCS) per unit volume [{\it Clothiaux et 
al.}, 1995]
is

\begin{equation} \eta(\vec r)=\int_{D} \sigma(D) {\cal N} (D,\vec r) dD.
\end{equation} Leaving aside technical factors, the power received at the radar
receiver antenna is proportional to $\eta(\vec r)$

\begin{equation} P(\vec r) \sim \frac{\eta(\vec r)}{r^2}. \end{equation} Due to
the particle movement, frequency shifts $w_m$, called Doppler shifts, appear in
the scattered signal

\begin{equation} w_m=-\frac{4\pi v_m}{\lambda_c}, \end{equation} 
where $m$ is the
index of the particle that moves with its so called fall velocity $v_m$ and
$\lambda_c$ is the wavelength of the sounding signal [{\it Clothiaux et al.},
1995].

The radar is equipped with signal processing units, which strip off the carrier
frequency and create signal voltages. By applying a Fourier transform over the
raw voltage time series one can obtain the power density $S_m$ at each velocity
$v_m$ [{\it Clothiaux et al.}, 1995]. From Eq. (3) one can express the power
density as a function of velocities rather than as a function of frequency. As
such, the power density $S_m$ at each velocity $v_m$ is directly related to the
total backscattering cross-section of all particles moving with a 
radial velocity
$v_m$ with respect to the radar. Let $\tilde S_m$ be the normalized 
power density
spectra at velocity $v_m$, i.e. \begin{equation} \tilde
S_m=\frac{S_m}{\sum_{1}^{{\cal N} } S_m}, \end{equation} where ${\cal N}$
corresponds to the number of particles in the resolution volume.

The radar signal processor units store the mean power-weighted radial velocity
within the resolution volume as

\begin{equation} <v> = \sum_{m=1}^{\cal N} v_m\tilde S_m, \end{equation}
which is usually referred to as the Doppler velocity.

The standard deviation of the power-weighted velocities about the mean is also
stored

\begin{equation} \sigma_v^2 = \sum_{m=1}^{\cal N} [v_m-<v>]^2 \tilde S_m
\end{equation} and is called the Doppler spectral width. These will 
be related to
physical properties later on, in Section III.

We study the signals of the backscattering cross section $\eta$, the Doppler
velocity $<v>$ and the Doppler spectral width $\sigma_v^2$ at different levels
within the vertical structure of cirrus clouds. We apply the detrended
fluctuation analysis (DFA) method [{\it Peng et al.}, 1994] to characterize the
correlations in these signals. The DFA method is suited to accurately 
quantifying
power-law correlations in noisy nonstationary signals with polynomial trends
[{\it Peng et al.}, 1994; {\it Vandewalle and Ausloos}, 1998; {\it Hu et al.},
2001; {\it Chen et al.}, 2002]. The advantage of the DFA method over 
conventional
methods, such as the power spectrum analysis, is that it avoids the spurious
detection of apparent long-range correlations that are an artifact of the
nonstationarity (related to linear and higher order polynomial trends in the
data). The DFA method has been tested at length by some of us on 
controlled time
series that contain long-range correlations superposed on a (nonstationary)
trend. Of note is a recent independent review of fractal fluctuation analysis
methods which determined that DFA is one of the most robust methods [{\it Taqqu
et al.}, 1995].

Briefly, the DFA method involves the following steps:

{\it (i)\/} The signal time series $\xi(i)$, $i=1,2,...,N$ is integrated, to
``mimic'' a random walk

\begin{equation} y(n)=\Sigma_{i=1}^n (\xi(i)-<\xi>), \end{equation} \noindent
where $<\xi>=\Sigma_{i=1}^N \xi(i)/N$.

{\it (ii)\/} The integrated time series is divided into boxes of equal length,
$\tau$.

{\it (iii)\/} In each box of length $\tau$, a least squares line is fit to the
data (representing the {\it trend\/} in that box). The $y$ coordinate of the
straight line segments is denoted by $z(n)$.

{\it (iv)\/} The integrated time series, $y(n)$, is detrended by 
subtracting the
local trend, $z(n)$, in each box.

{\it (v)\/} For a given box size $\tau$, the characteristic size of fluctuation
$F(\tau)$ for this integrated and detrended time series is calculated:

\begin{equation} F(\tau) = \sqrt{{1 \over \tau } {\sum_{n=k\tau+1}^{(k+1)\tau}
{\left[y(n)- z(n)\right]}^2}} \qquad 
k=0,1,2,\dots,\left(\frac{N}{\tau}-1\right)
\end{equation}

{\it (vi)\/} The above computation is repeated over all time scales (box sizes
$\tau$) to provide a relationship between $F(\tau)$ and the box size 
$\tau$ (i.e.
the size of the window of observation).

A power law relation between the average root-mean-square fluctuation function
$F(\tau)$ and the size of the observation window indicates the presence of so
called scaling [{\it Ma}, 1976], i.e. when the fluctuation correlations can be
characterized by a scaling exponent $\alpha$, the self-similarity parameter, as
in

\begin{equation} F(\tau) \sim \tau^{\alpha}. \end{equation}

For uncorrelated random walk fluctuations the $F(\tau)$ function scales with an
exponent $\alpha=1.5$ [{\it Schroeder}, 1991; {\it Addison}, 1997, {\it
Turcotte}, 1997]. An exponent $\alpha>1.5$ indicates a persistent correlated
behavior in the fluctuations, while an exponent $\alpha <1.5$ characterizes
anticorrelations. In other words, if $\alpha <1.5$, the signal 
increments  during
the nonoverlapping successive time intervals of size $\tau$ tend to be of
opposite signs, so that the signal $y(n)$  has a tendency to decrease 
in the next
time step if it has had an increasing tendency in the previous one and vice
versa. The feature is called antipersistency, or there are 
$anticorrelations$. If
$\alpha>1.5$ the signal increments tend to have the same signs, so that $y(n)$
tends to increase in the future if it has had an increasing tendency 
in the past,
and conversely. The feature is called $persistency$. Physically, a persistent
system is going to increase its statistical mean with time as if a positive
feedback dominates the system. The antipersistency expresses a tendency of the
values of increments to compensate for each over; this prevents the 
process from
blowing up or going down very fast. Such a system tends to eliminate 
deviations,
as if there is  a negative feedback. An antipersistent time series visit, on
average, the mean value more often than an ordinary one; the signal is said to
present $anticorrelations$. Notice that the scaling exponent $\alpha$ can be
related to the exponent $\beta$ of the power spectral density $S(f)\sim
f^{-\beta}$ of such a time series $y(t)$, i.e. $\beta=2\alpha-1$ [{\it Heneghan
and McDarby}, 2000].

There are different ways of detrending the signal: one can fit the data in a
window to a linear, quadratic, cubic [{\it Vandewalle and Ausloos}, 1998] or
higher degree polynomial and then study the fluctuations of the signal.
Discussions on different ways of detrending a signal, including a periodic and
power-law trends, can be found in {\it Hu et al.} [2001].

\section{Data and Data Analysis}

We analyze data collected with a millimeter wave cloud radar that operates at
34.86~GHz $\pm$ 
200~MHz.\footnote{http://www.arm.gov/docs/sites/sgp/sgp.html} The
radar produce measurements at four modes [{\it Clothiaux et al.}, 
2000] that are
used to obtain a best estimate to cloud signal following an interpolation
procedure described in [{\it Clothiaux et al.}, 2000]. The radar measures the
backscattering cross-section of the emitted wave per unit volume of cloud
particles. Assuming that the Rayleigh approximation is valid and ice 
crystals are
spherical particles (we are aware of the approximation) this cross section is
proportional to the sixth moment of the particle size and also 
depends on its ice
water content [{\it Mace}, 1997]. The logarithm of the backscattering cross
section is called the radar reflectivity and is the quantity typically used in
the field. However, because we aim at studying the fluctuations, we 
prefer to use
the raw backscattering cross section signal.

Continuous measurements of (i) the backscattering cross section per unit volume
$\eta$, (ii) the Doppler velocity $<v>$ and (iii) the Doppler spectral width
$\sigma_v^2$ are recorded with 10~s temporal resolution 24 hours a 
day and with a
45~m spatial vertical resolution at 512 height levels. Radar reflectivity data
measured with an accuracy of 0.5~dB are shown in Fig. 1.

An order of magnitude value for the horizontal velocity at which such 
clouds move
is ~10 m/s. Therefore the volume examined at each data point can be 
considered to
reflect the local ice content and velocity fluctuation in a fixed volume, as if
the cloud is locally stable. The number of particles in the volume is 
assumed to
be a constant during the measurements. If this is not the case, from a
statistical physics point of view, one should work in a grand 
canonical ensemble
formalism and subsequently introduce another {\it ad hoc} model than in  [{\it
Clothiaux et al.}, 2000].

Studies along isotherms or isobars would be of great interest. This information
being missing, we have decided to cut the cloud thickness into isodepths,
identifying the cloud as being between its top and its bottom layer, i.e. at
which the signal level is above -45~dB. We study the time series of the three
sets of measurements at certain isodepths (relative to the thickness $h$ of the
cloud), i.e. top, 0.75~$h$, 0.5~$h$, 0.25~$h$, and bottom. These profiles are
represented in Figs. 2 and 3, from which the cloud height, thickness and their
time evolution can be read.

In this study in order $not$ to introduce spurious time correlation effects, we
avoid large data gaps within the cloud life time, e.g. the gaps 
observed between
9:00 and 18:00 GMT on Jan. 26, 1997. Thus we consider that the time extent of
this $winter$ cloud at this location was between 18:00 GMT on Jan. 26 and 06:00
GMT on Jan. 27, 1997, which amounts to 4315 data points. The $fall$ 
cloud extends
from 18:00 to 24:00 GMT on Sept. 26, 1997 which amounts to 2048 data points.

The time series of the backscattering cross section $\eta$, Doppler velocity
$<v>$, and Doppler spectral width $\sigma_v^2$ at the specified heights (or
isodepths) within the winter cirrus cloud are plotted in Fig. 4(a-c). 
Results for
each DFA-function $F(\tau)$ are plotted in Fig. 5(a-e). The DFA-functions are
grouped depending on the profile level, i.e. they refer to the top, 0.75~$h$,
0.5~$h$, 0.25~$h$, or the bottom of the cloud. For the fall cirrus cloud
(observed at the same experimental site), the data series of the 
three quantities
of interest have been also analyzed along the same profile levels as the winter
cloud (see Fig. 6). The results from the DFA analysis at the different height
levels are displayed in Fig.7(a-e).

The scaling analysis has been used performing statistical tests to 
find the best
slopes for the fits, and the corresponding scaling regimes. Each 
scaling behavior
is characterized by slightly different scaling exponents, $\alpha_1$ and
$\alpha_2$ at low and large time lag $\tau$ respectively; we find a 
crossover at
$\tau_x \simeq 3-5$~min for all data series and physical quantities. The values
of the $\alpha$-exponents for the two scaling regimes and for the three studied
quantities are summarized in Table 1 and Table 2 for the winter cloud and the
fall cloud. Their values, whence the clouds, can be compared with each other,
from signal type to signal type, and on both sides of the crossover 
time lag for
a given signal. The error of the estimate of each $\alpha$ value is thus the
error of the slope and is defined as the square root of residual sum of squares
divided by the number of points in the fit minus two and by the sum 
of squares of
the deviation with respect to the mean of the time lags [{\it Bowman and
Robinson}, 1987]. The correlation coefficient for all estimates is in the range
between 0.992 and 0.998.

We find that the $\alpha_1$ values have a maximum in the center of 
the cloud for
both cloud cases. We note an increase of $\alpha_1$ from bottom to top for the
Doppler velocity of the winter cloud, while the value of $\alpha_1$ is rather
constant for the fall cloud. The $\alpha_1$ values remain constant for the
$\eta$, $<v>$ and $\sigma_v^2$ data and for both cloud cases, and 
close to $1.5$,
except for $<v>$ at the cloud bottom in the Jan. 26 and 27, 1997
case.\footnote{The short-range correlations may be influenced by the
interpolation procedure that produces the best estimate to cloud 
signal from the
four modes measurements of the millimeter wave radar. The current interpolating
procedure is complicated [{\it Clothiaux et al.}, 2000] and mimics the signal
from the four modes in a systematic but non-uniform way.}

The $\alpha_2$ values are systematically close to or below $1.0$, for $<v>$ and
$\sigma_v^2$. The backscattering cross section $\eta$ values fall around $1.0$.
The $\alpha_2$ values decrease from bottom to top for $<v>$ and 
$\sigma_v^2$, but
present a maximum in the fall cloud center for the $\eta$ data. The $\alpha_2$
values reach a smooth maximum in the center of the cloud for both $<v>$ and
$\sigma_v^2$ data. The $\alpha_2$ values reach a maximum in the center of the
cloud for the $\eta$, $<v>$ and $\sigma_v^2$ data and so in both cloud cases,
though all these values are quite close to $1.0$.

Thus, scaling exponents at the top of the cloud seem to have usually a lower
value as compared to that in the bulk of the cloud, - {\it except for $<v>$}.
This could be due to a physical effect or arise from the instrumental noise
contribution: at such heights the real signal may be low as compared to the
instrumental
noise.\footnote{$http://www.arm.gov/docs/instruments/static/mmcr.html$}

To check the robustness of the results it is essential to perform a test on
surrogate data [{\it Schreiber}, 1999; {\it Schreiber and Schmitz}, 2000]: we
randomly shuffled the amplitudes of (i) the backscattering cross 
section per unit
volume $\eta$, (ii) the Doppler velocity $<v>$ and (iii) the Doppler spectral
width $\sigma_v^2$ signals for winter and fall clouds. It is known [{\it
Viswanathan et al.}, 2002] that the fat tailed distributions are thought to be
caused by long-range volatility correlations. Destroying all correlations by
shuffling the order of the fluctuations, is known to cause the fat tails almost
to vanish. We have found that the long-range correlations do vanish 
in surrogate
data as seen from the DFA-functions plotted in Figs. 8 and 9 and 
Tables 3 and 4.
This, together with an excellent systematics in the root mean square linear fit
correlation coefficients, leads to put substance to the results found 
here above.

\section{Discussion}

Four main points seem to need some emphasis and bring newly 
interesting results:

(I) We find that DFA-functions $F(\tau)$ for all radar backscattering 
time series
exhibit a crossover at about 4~min. We observe a $different$ $scaling$ behavior
for the backscattering cross section $\eta$, the Doppler velocity $<v>$ and the
Doppler spectral width $\sigma_v^2$ between the top of the cloud from the bulk,
and from the bottom, especially for large time scales. Short time range
($<4$~min) scaling of $\eta$ is {\it completely uncorrelated} with
$\alpha_1\approx 1.5$. Long-range correlations of $\eta$, $<v>$ and 
$\sigma_v^2$
are quite similar (within error bars) for the bulk of the cloud with $\alpha_2
\approx 0.90$ for $<v>$ and $\sigma_v^2$ and $\alpha_2\approx 1.00$ for $\eta$.
Such $\alpha$-exponents (or $\beta=1$) characterize a special 
physical phenomenon
of interest [{\it Schroeder}, 1991], that is $1/f$ noise [{\it 
Shlesinger}, 1987;
{\it Weismann}, 1988; {\it Shlesinger and West}, 1988; {\it West and 
Shlesinger},
1989].

(II) Cirrus cloud temperature measurements with radiosondes show that the
temperature within the cloud field usually varies smoothly between about
$-43^{\circ}$~C and $-33^{\circ}$~C. Spontaneous freezing is thought 
to occur at
temperatures colder than about $-40^{\circ}$~C. Assuming therefore that
kinetically driven {\it homogeneous nucleation} is the primary formation
mechanism at these temperatures [{\it Kiang et. al}, 1971; {\it 
Sassen and Dodd},
1988; {\it Tabazadeh et al.}, 1997a, 1997b] supercooled liquid water drops are
thus seen to nucleate (and die rapidly) in water-satured updrafts.

(III) However beside the $\alpha=1.5$ value, i.e. indicating no correlation at
all in the signal fluctuations at short time ranges, thus suggesting 
a birth and
death (nucleation) process, the scaling difference at the crossover, to a $1/f$
process, suggests propagation effects in the cirrus cloud at long time scales,
similar to growth and percolation features. The Doppler velocity 
measurements of
the cirrus ice particles shown in Fig. 2b suggest $updrafts$ in fact, ...
therefore updrafts serve as mechanisms for the aggregation of ice particles at
long time ranges. The data analysis is consistent with (and somewhat 
proves) such
an intuitive physical mechanism.

Further physical insight is obtained if we translate the $\eta$, $<v>$ and
$\sigma_v^2$ data from scattering information into usual fluid mechanics
parameters. Even though the data is analyzed through a model with 
constant number
of particles with a given (spherical) shape, and other standard approximations,
nevertheless the backscattering cross section is a measure of the scattering
intensity, thus clearly a measure of the ice content, in principle including
particle shape, size and distribution. Therefore the $\eta$-DFA function is the
direct signature of the {\it density-density correlation function}, i.e. the
zero-wave vector structure factor. In other words, it is a measure of the cloud
(fluid) {\it compressibility} or bulk modulus [{\it Chaikin and 
Lubensky}, 1995].
The $<v>$ data leads to information about the particle velocity in the cloud,
while the $<v>$-DFA function relates to the {\it velocity-velocity correlation
function}. Even assuming a constant density, or a constant and 
identical mass for
each particles,  (... we are aware that this is an approximation), $<v>$-DFA
measures the correlations in particle momenta and thus relates to the 
$viscosity$
[{\it Huang}, 1967] of the inner cloud structure. On the contrary, $if$ the
particle density is not homogeneous the $<v>$-DFA function still has a physical
meaning, i.e. it is a signature for the $diffusivity$ or diffusion coefficient
[{\it Chaikin and Lubensky}, 1995]. Finally the $\sigma_v^2$ data is 
a measure of
the dissipated power due to inelastic scattering, through the velocity square
[{\it Huang}, 1967], and is thus related to the (cloud) temperature. The
$\sigma_v^2$-DFA function is thus a measure of the {\it temperature-temperature
correlation function}, whence the {\it thermal conductivity} [{\it 
Huang}, 1967].

In principle, the equations of motion for the above three physical quantities,
i.e. ice density, ice crystal velocity and temperature in such cirrus 
clouds can
be written as Navier-Stokes equations, - in terms of these three basic physical
coefficients.  They constitute an essential knowledge to be acquired before
meaningful physical modelling. Moreover, the diffusion coefficient and the
viscosity coefficient can serve as inputs in the corresponding microscopic
evolution equations, i.e. the Langevin equations, with appropriate noise and
drift terms.

In conclusion, the time dependence characteristics of physical quantities in
cirrus clouds have been obtained. Short time range correlations of radar back
scattering cross section are uncorrelated, i.e. are Brownian-like; at time lags
longer than about 3~ to 5~ min correlations are of the $1/f$ noise 
type. Certain
asymmetry is seen to exist from cloud top to bottom in respective scaling
properties. This can be understood to be due to different mechanisms on the
boundaries and in the bulk of the cloud, depending on the time scales. The data
analysis leads to a consistent physical picture on the inner nonequilibrium
structure, i.e. ice crystal nucleation and growth. Finally, the physical
description of such cirrus clouds, is shown to be related to the bulk modulus,
viscosity and thermal conductivity.

\vskip 1cm

{\noindent \bf \large Acknowledgements} \vskip 0.6cm KI and HES thank NATO CLG
976148. PChI and HES thank NIH/National Center for Research Resources (Grant
No.~P41RR13622) and NSF for support. This research is supported in part by the
Department of Energy through grant Battelle 327421-A-N4. We acknowledge using
data collected at the Southern Great Plains site of Atmospheric Radiation
Measurements (ARM) program.

\vskip 1cm

\begin{center} {\bf REFERENCES} \end{center}

Addison, P.S., {\it Fractals and Chaos}, Inst. of Phys., Bristol, 1997.

\vspace*{0.4cm}

Ashkenazy, Y., P. Ch. Ivanov, S. Havlin, C.-K. Peng, A. L. 
Goldberger, and H. E.
Stanley, Magnitude and sign correlations in heartbeat fluctuations, {\it Phys.
Rev. Lett., 86}, 1900-1903, 2001.

\vspace*{0.4cm}

Atlas, D., S.Y. Matrosov, A.J. Heymsfield, M.-D. Chou, and D.B. 
Wolff, Radar and
radiation properties of ice clouds, {\it J. Appl. Meteor., 34}, 
2329-2345, 1995.

\vspace*{0.4cm}

Ausloos, M., in: {\it Vom Billardtisch bis Monte Carlo - Spielfelder der
Statistischen Physik}, K.H. Hoffmann and M. Schreiber, Eds. (Springer, Berlin,
2001) pp. 153-168.

\vspace*{0.4cm}

Ausloos, M., and K. Ivanova, Precise (m,k)-Zipf diagram analysis of 
mathematical
and financial time series when m=6 and k=2, {\it Physica A, 270}, 
526-542, 1999.

\vspace*{0.4cm}

Ausloos, M., and K. Ivanova, Power law correlations in the Southern Oscilation
Index fluctuations characterizing El Nino, {\it Phys. Rev. E, 63}, 
047201 (1-4),
2001.

\vspace*{0.4cm}

Bowman, A.W. and D.R. Robinson, {\it Introduction to statistics}, Institute of
Physics Publishing, Bristol, 1987.

\vspace*{0.4cm}

Brockwell, P.J., and R.A. Davis, {\it Introduction to Time Series and
Forecasting}, Springer Text in Statistics, Springer, Berlin, 1998.

\vspace*{0.4cm}

Bunde, A., S. Havlin, E. Koscielny-Bunde, and H.-J.Schellnhuber, Long term
persistence in the atmosphere: global laws and tests of climate models, {\it
Physica A, 302}, 255-267, 2001.

\vspace*{0.4cm}

Chaikin, P.M., and T.C. Lubensky, {\it Principles of Condensed Matter Physics},
Cambridge U.P., Cambridge, 1995.

\vspace*{0.4cm}

Chen, Z., P. Ch. Ivanov, K. Hu, and H.E. Stanley, Effect of 
nonstationarities on
detrended fluctuation analysis, {\it Phys. Rev E, 65}, 041107 (1-15) 2002.

\vspace*{0.4cm}

Clothiaux, E.E., T.P. Ackerman, R.T. Marchand, J. Verlinde, D.M. Babb, and C.S.
Ruf, in {\it Proc. of the NATO ASI on Remote Sensing of Processes Governing
Energy and Water Cycles in the Climate System}, Ploen, Germany, May 1-12, 1995.

\vspace*{0.4cm}

Clothiaux, E.E., T.P. Ackerman, G.G. Mace, K.P. Moran, R.T. Marchand, M.A.
Miller, and B.E. Martner, Objective determination of cloud heights and radar
reflectivities using a combination of active remote sensors at the ARM CART
sites, {\it J. Appl. Meteor., 39}, 645-665, 2000.

\vspace*{0.4cm}

Davis, A., A. Marshak, W. Wiscombe, and R. Cahalan, Scale Invariance of liquid
water distributions in marine stratocumulus. Part I: Spectral properties and
stationarity issues, {\it J. Atmos. Sci., 53}, 1538-1558, 1996.

\vspace*{0.4cm}

Gingle, A., and T.M. Knasel, Undergraduate laboratory investigation of the
dielectric constant of ice, {\it Amer. J. Phys., 43}, 161-167, 1975.

\vspace*{0.4cm}

Hausdorff, J.M., S.L. Mitchell, R. Firtion, C.-K. Peng, M. Cudkowicz, 
J. Y. Wei,
and A.L. Goldberger, Altered fractal dynamics of gait: reduced stride-interval
correlations with aging and Huntington's disease, {\it J. Appl. Physiol., 82},
262-269, 1997.

\vspace*{0.4cm}

Heneghan, C., and G. McDarby, Establishing the relation between detrended
flulctuation analysis and power spectral density analysis for stochastic
processes, {\it Phys. Rev. E, 62}, 6103-6110, 2000.

62 6103 2000 \vspace*{0.4cm}

Heymsfield, A.J., and C.M.R. Platt, A parameterization of the particle size
spectrum of ice clouds in terms of the ambient temperature and the ice water
content, {\it J. Atmos. Sci., 41}, 846-855, 1984.

\vspace*{0.4cm}

Hu, K., Z. Chen, P. Ch. Ivanov, P. Carpena, and H.E. Stanley, Effect 
of trends on
detrended fluctuation analysis, {\it Phys. Rev E, 64}, 011114 (1-19) (2001).

\vspace*{0.4cm}

Huang, K., {\it Statistical Mechanics}, J. Wiley, New York, 1967.

\vspace*{0.4cm}

Huckaby, D.A., R. Pitis, A.K. Belkasri, M. Shinmi, and L. Blum, 
Molecular mirror
images, dense ice phases, organic salts at interfaces, and electrochemical
deposition: exotic applications of the Pirogov-Sinai theory, {\it Physica A,
285}, 211-219, 2000.

\vspace*{0.4cm}

Ivanov, P.~Ch., M.~G.~Rosenblum, C.-K.~Peng, J.~E.~Mietus, S.~Havlin, H.~E.
Stanley, and A.~L.~Goldberger, Scaling behaviour of heartbeat 
intervals obtained
by wavelet-based time-series analysis, {\it Nature, 383}, 323-327, 1996.

\vspace*{0.4cm}

Ivanov, P.~Ch., A. Bunde, L. A. N. Amaral, S. Havlin, J. Fritsch-Yelle, R. M.
Baevsky, H. E. Stanley, and A. L. Goldberger, Sleep-wake difference in scaling
properties of the human heart, {\it Europhys. Lett., 48}, 594-600, 1999.

\vspace*{0.4cm}

Ivanova, K., M. Ausloos, E.E. Clothiaux, and T.P. Ackerman, Break-up of stratus
cloud structure predicted from non-Brownian motion liquid water and brightness
temperature fluctuations, {\it Europhys. Lett., 52}, 40-46, 2000.

\vspace*{0.4cm}

Ivanova, K., E.E. Clothiaux, H.N. Shirer, T.P. Ackerman, J.C. Liljegren, and M.
Ausloos, Evaluating the quality of ground-based microwave radiometer 
measurements
and retrievals using detrended fluctuation and spectral analysis 
methods, {\it J.
Appl. Meteor., 41}, 56-68, 2002.

\vspace*{0.4cm}

Kantz, H., and Th. Schreiber, {\it Nonlinear Time Series Analysis}, Cambridge
U.P., Cambridge, 1997.

\vspace*{0.4cm}

Kiang, C.S., D. Stauffer, G.H. Walker, O.P. Puri, J.D. Wise, Jr., 
E.M. Patterson,
A reexamination of homogeneous nucleation theory, {\it J. Atmos. Sci., 28},
1222-1232, 1971.

\vspace*{0.4cm}

Koscielny-Bunde, E., A. Bunde, S. Havlin, H. E. Roman, Y. Goldreich, and H.-J.
Schellnhuber, Indication of a Universal Persistence Law Governing Atmospheric
Variability, {\it Phys. Rev. Lett., 81}, 729-732, 1998.

\vspace*{0.4cm}

Koscielny-Bunde, E., A. Bunde, S. Havlin, and Y. Goldreich, Analysis of daily
temperature fluctuations, {\it Physica A, 231}, 393-396, 1993.

\vspace*{0.4cm}

Liou, K. N., S. C. Ou, Y. Takano, J. Roskovensky, G.C. Mace, K. Sassen, and M.
Poellot,  Remote sensing of three-dimensional inhomogeneous cirrus clouds
using satellite and mm-wave cloud radar data, {\it Geophys. Res. Lett., 29},
14846-14857, 2002.

\vspace*{0.4cm}

Ma S.K., {\it Modern Theory of Critical Phenomena}, Benjamin, 
Reading, MA, 1976.

\vspace*{0.4cm}

Mace, G.G., T.P. Ackerman, E.E. Clothiaux, and B.A. Albrecht, A study of
composite cirrus morphology using data from a 94-GHz radar and 
correlations with
temperature and large scale vertical motion, {\it J. Geophys. Res., 102},
13~581-13~593, 1997.

\vspace*{0.4cm}

Mace, G.G., T.P. Ackerman, P. Minnis, and D.F. Young, Cirrus layer 
microphysical
properties derived from surface based millimeter radar and infrared radiometer
data, {\it J. Geophys. Res., 103}, 23~207-23~216, 1998.

\vspace*{0.4cm}

Malamud, B.D. and D.L. Turcotte, Self-affine time series: measures of weak and
strong persistence {\it J. Stat. Plann. Infer., 80}, 173-196, 1999.

\vspace*{0.4cm}

Mantegna, R.N., and H.E. Stanley, Scaling behavior in the dynamics of 
an economic
index, {\it Nature, 376}, 46-49, 1995.

\vspace*{0.4cm}

Maurellis, A., Could water vapour be the culprit in global warming?, 
{\it Physics
World, 14}, 22-23, 2001.

\vspace*{0.4cm}

McFarquhar, G.M., and A.J. Heymsfield, Parameterization of tropical cirrus ice
crystal size distributions and implications for radiative transfer: 
Results from
CEPEX, {\it J. Atmos. Sci., 54}, 2187-2200, 1997.

\vspace*{0.4cm}

Mishima, O., and H. E. Stanley, Decompression-induced melting of ice IV and the
liquid-iquid transition in water, {\it Nature, 392}, 164-167, 1998.

\vspace*{0.4cm}

Mercik, S., Z.~Siwy, and K.~Weron, What can be learnt from the 
analysis of short
time series of ion channel recordings, {\it Physica A, 276}, 376-390, 2000.

\vspace*{0.4cm}

Peng, C.-K., S.V. Buldyrev, S. Havlin, M. Simons, H.E. Stanley, and A.L.
Goldberger, Mosaic organization of DNA nucleotides, {\it Phys. Rev. E, 49},
1685-1689, 1994.

\vspace*{0.4cm}

Peters, O., Ch. Hertlein, and K. Christensen, A Complexity View of 
Rainfall, {\it
Phys. Rev. Lett., 88}, 018701 (1-4), 2002.

\vspace*{0.4cm}

Petrenko, V.F., and R.W. Whitworth, {\it Physics of Ice}, Oxford University
Press, New York, 1999.

\vspace*{0.4cm}

Putrino, A. and M. Parrinello, Anharmonic Raman Spectra in 
High-Pressure Ice from
Ab Initio Simulations, {\it Phys. Rev. Lett., 88}, 176401 (1-4), 2002.

\vspace*{0.4cm}

Rogers, R.R., {\it Short Course in Cloud Physics}, Pergamon Press, New York,
1976.

\vspace*{0.4cm}

Rosenfeld, D., and W. Woodley, Pollution and clouds, {\it Physics World, 14},
33-37, 2001.

\vspace*{0.4cm}

Sassen, K., and G.C. Dodd, Homogeneous nucleation rate for highly supercooled
cirrus cloud droplets, {\it J. Atmos. Sci., 45}, 1357-1369, 1988.

\vspace*{0.4cm}

Schreiber, Th., Interdisciplinary application of nonlinear time series methods,
{\it Phys. Rep., 308}, 1-64, 1999.

\vspace*{0.4cm}

Schreiber, Th., and A. Schmiltz, Surrogate time series, {\it Physica D, 142},
346-382, 2000.

\vspace*{0.4cm}

Schroeder, M., {\it Fractals, Chaos and Power Laws}, W.H. Freeman and Co., New
York, 1991.

\vspace*{0.4cm}

Shlesinger, M.F., Fractal time and $1/f$ noise in complex systems. {\it Ann. NY
Acad. Sci., 504}, 214-228, 1987.

\vspace*{0.4cm}

Shlesinger, M.F., and B. J. West, $1/f$ noise versus $1/f^{\beta}$ noise. in:
{\it Random Fluctuations and Pattern Growth: Experiments and Models}, 
eds. H. E.
Stanley and N. Ostrowsky, Kluver Academics, Boston, 1988.

\vspace*{0.4cm}

Smith, S.A., and A.D. Del Genio, Analysis of aircraft, radiosonde and radar
observations in cirrus clouds observed during FIRE II, {\it J. Atmos. 
Sci., 58},
451-461, 2001.

\vspace*{0.4cm}

Stanley, H.E., S.V. Buldyrev, A.L. Goldberger, S. Havlin, C.-K. Peng, and M.
Simons, Long-range power-law correlations in condensed matter physics and
biophysics, {\it Physica A, 200}, 4-24, 1993.

\vspace*{0.4cm}

Stephens, G.L., S.-C. Tsay, P.W. Stackhouse, Jr., and P.J. Flatau, 
The relevance
of the microphysical and radiative properties of cirrus clouds to climate and
climatic feedback, {\it J. Atmos. Sci., 47}, 1742-1754, 1990.

\vspace*{0.4cm}

Sundqvist, H.E., E. Berge, and J.E. Kristiansson, Condensation and cloud
parameterization studies with a mesoscale numerical weather prediction model,
{\it Mon. Wea. Rev., 117}, 1641-1657, 1989.

\vspace*{0.4cm}

Tabazadeh, A., E.J. Jensen, and O.B. Toon, A model description for cirrus cloud
nucleation from homogeneous freezing of sulfate aerosols, {\it J. 
Geophys. Res.,
102}, 23~2845-23~850, 1997.

\vspace*{0.4cm}

Tabazadeh, A., O.B. Toon, and E.J. Jensen, Formation and implications of ice
particle nucleation in the stratosphere, {\it Geophys. Res. Lett., 24},
2007-2010, 1997.

\vspace*{0.4cm}

Taqqu, M.S., V. Teverovsky, and W. Willinger, Estimators for long-range
dependence: an empirical study, {\it Fractals, 3}, 785-798, 1995.

\vspace*{0.4cm}

Turcotte, D.L., {\it Fractals and Chaos in Geology and Geophysics}, Cambridge
U.P., Cambridge, 1997.

\vspace*{0.4cm}

Vandewalle, N., and M. Ausloos, Coherent and random sequences in financial
fluctuations, {\it Physica A, 246}, 454-459, 1997.

\vspace*{0.4cm}

Vandewalle, N., and M. Ausloos, Extended detrended fluctuation analysis for
financial data,{\it Int. J. Comput. Anticipat. Syst., 1}, 342-349, 1998.

\vspace*{0.4cm}

Viswanathan, G. M., U. L. Fulco, M. L. Lyra and M. Serva, The origin of fat
tailed distributions in financial time series, (arXiv: cond- mat/ 0112484 v3 3
Jan 2002)

\vspace*{0.4cm}

Wallace, J.M., and P.V. Hobbs, {\it Atmospheric Science}, Academic Press, New
York, 1977.

\vspace*{0.4cm}

Weismann, M., $1/f$ noise and other slow, nonexponential kinetics in condensed
matter, {\it Rev. Mod. Phys., 60}, 537-571, 1988.

\vspace*{0.4cm}

Weiss, J., and J.R. Grasso, Acoustic emission in single crystals of 
ice, {\it J.
Phys. Chem. B, 101}, 6113-6117, 1997.

\vspace*{0.4cm}

West, B.J., and M.F. Shlesinger, On the ubiquity of $1/f$ noise, {\it Int. J.
Mod. Phys. B, 3}, 795-819, 1989.

\vspace*{0.4cm}

Zhang, G.J., and N.A. McFarlane, Sensitivity of climate simulations to the
parameterization of cumulus convection in the Canadian Climate Centre general
circulation model, {\it Atmos.-Ocean., 33}, 407-446, 1995.

\vspace*{0.4cm}

Zurovac-Jevtic, D., Development of a cirrus parameterization scheme: 
Performance
studies in HIRLAM, {\it Mon. Wea. Rev., 127}, 470-485, 1999.

{\large \bf Figure Captions}

\vskip 0.5cm {\bf Figure 1} -- Radar reflectivity observations on (a) Jan. 26,
(b) 27, 1997 and (c) on Sept. 26, 1997 at the Southern Great Plains site of
Atmospheric Radiation Measurements program.

\vskip 0.5cm {\bf Figure 2} -- Cloud profiles at top and bottom and at $0.75$,
$0.5$ and $0.25$ of the thickness $h$ of the $winter$ cirrus cloud (Fig.1a,b)
along which the backscattering cross section, Doppler velocity and Doppler
spectral width are studied.

\vskip 0.5cm {\bf Figure 3} -- Cloud profiles at top and bottom and at $0.75$,
$0.5$ and $0.25$ of the thickness $h$ of the $fall$ cirrus cloud 
along which the
backscattering cross section, Doppler velocity and Doppler spectral width are
studied. Measurements are obtained at the Southern Great Plains site of
Atmospheric Radiation Measurements program on Sept. 26, 1997 (Fig. 1c).

\vskip 0.5cm {\bf Figure 4} -- (a) Backscattering cross section per unit volume
at different relative heights to the thickness $h$ of the cirrus cloud as
measured on Jan. 26 and 27, 1997 at the Southern Great Plains site of 
Atmospheric
Radiation Measurements program (data in Fig. 1a,b). (b) Doppler velocity (m/s).
The positive spike between 20 and 22 h of the "bottom" data series is 
of order of
11~m/s.(not shown) at heights in the cloud shown in Fig. 2.; (c) 
Doppler spectral
width (m/s) at heights in the cloud shown in Fig. 2.

\vskip 0.5cm {\bf Figure 5} -- DFA-functions for backscattering cross section
(circles), Doppler velocity (triangles) and Doppler spectral width (squares) at
the (a) top, (b) 0.75, (c) 0.50, (d) 0.25 relative to the thickness $h$ of the
cloud, and (e) bottom of the $winter$ cloud shown in Fig. 2. DFA-functions are
displaced for readability. The $\alpha$-values are listed in Table 1.

\vskip 0.5cm {\bf Figure 6} -- (a) Backscattering cross section per unit volume
at different relative heights to the thickness $h$ of the cirrus cloud as
measured on Sept. 26, 1997 at the Southern Great Plains site of Atmospheric
Radiation Measurements program (Fig. 1c). (b) Doppler velocity (m/s) at heights
in the cloud shown in Fig. 3; (c) Doppler spectral width (m/s) at different
heights in the cloud shown in Fig. 3.

\vskip 0.5cm {\bf Figure 7} -- DFA-functions for backscattering cross section
(circles), Doppler velocity (triangles) and Doppler spectral width (squares) at
the (a) top, (b) 0.75, (c) 0.50, (d) 0.25 relative to the thickness $h$ of the
cloud, and (e) bottom of the cloud shown in Fig. 3. DFA-functions are displaced
for readability. The $\alpha$-values are listed in Table 2.

\vskip 0.5cm {\bf Figure 8} -- DFA-functions for shuffled data of: 
backscattering
cross section (circles), Doppler velocity (triangles) and Doppler 
spectral width
(squares) at the (a) top, (b) 0.75, (c) 0.50, (d) 0.25 relative to 
the thickness
$h$ of the cloud, and (e) bottom of the $winter$ cloud shown in Fig. 2.
DFA-functions are displaced for readability. The $\alpha$-values are listed in
Table 3.

\vskip 0.5cm {\bf Figure 9} -- DFA-functions for the shuffled data of:
backscattering cross section (circles), Doppler velocity (triangles) 
and Doppler
spectral width (squares) at the (a) top, (b) 0.75, (c) 0.50, (d) 0.25 
relative to
the thickness $h$ of the cloud, and (e) bottom of the $fall$ cloud 
shown in Fig.
3. DFA-functions are displaced for readability. The $\alpha$-values 
are listed in
Table 4.

\newpage

\begin{table}[ht]

\caption{Values of the $\alpha_1$ and $\alpha_2$-exponents from the 
DFA analysis
of the backscattered cross section $\eta$, Doppler velocity $<v>$ and Doppler
spectral width $\sigma_v^2$ data shown in Fig. 4(a-c); $winter$ cloud. The
crossover $\alpha_1$ to $\alpha_2$ is at about 3~min. The error of the estimate
of each $\alpha$ value is thus the error of the slope and is defined as the
square root of residual sum of squares divided by the number of 
points in the fit
minus two and by the sum of squares of the deviation with respect to 
the mean of
the time lags [{\it Bowman and Robinson}, 1987].}
\tabcolsep=0pt
\begin{center} \begin{tabular}{||c|c|c||c|c||c|c||} \hline position &
\multicolumn{2}{c||}{$\eta$} & \multicolumn{2}{c||}{$<v>$} &
\multicolumn{2}{c||}{$\sigma_v^2$ } \\ \hline &$\alpha_1$ &$\alpha_2$ 
&$\alpha_1$
&$\alpha_2$ &$\alpha_1$ &$\alpha_2$ \\ \hline top &$1.36\pm0.03$ & 
$0.89\pm0.01$
& $1.43\pm0.07$ & $0.76\pm0.01$ & $1.24\pm0.04$ & $0.73\pm0.01$\\ 0.75 h
&$1.41\pm0.04$ & $1.11\pm0.01$ & $1.38\pm0.04$ & $0.88\pm0.01$ & 
$1.36\pm0.03$ &
$0.90\pm0.02$\\ 0.50 h &$1.62\pm0.10$ & $0.94\pm0.01$ & $1.42\pm0.04$ &
$0.96\pm0.01$ &$1.35\pm0.04$ & $0.92\pm0.02$\\ 0.25 h &$1.43\pm0.04$ &
$1.15\pm0.02$ & $1.29\pm0.03$ & $0.93\pm0.01$ &$1.31\pm0.04$ & $0.90\pm0.02$\\
bottom &$1.46\pm0.09$ & $1.27\pm0.03$ & $1.09\pm0.05$ & $0.70\pm0.02$ &
$1.38\pm0.05$& $1.02\pm0.01$\\ \hline \end{tabular} \end{center} \end{table}

\begin{table}[ht]

\caption{Values of the $\alpha_1$ and $\alpha_2$-exponents from the 
DFA analysis
of backscattering cross section $\eta$, Doppler velocity $<v>$ and Doppler
spectral width $\sigma_v^2$ data shown in Fig. 6(a-c); $fall$ cloud. 
A crossover
is observed at about 5~min.}
\tabcolsep=0pt

\begin{center} \begin{tabular}{||c|c|c||c|c||c|c||} \hline position &
\multicolumn{2}{c||}{$\eta$} & \multicolumn{2}{c||}{$<v>$} &
\multicolumn{2}{c||}{$\sigma_v^2$ } \\ \hline &$\alpha_1$ &$\alpha_2$ 
&$\alpha_1$
&$\alpha_2$ &$\alpha_1$ &$\alpha_2$ \\ \hline top &$1.32\pm0.05$ & 
$0.66\pm0.04$
& $1.34\pm0.06$ & $0.66\pm0.01$ & $1.32\pm0.03$ & $0.69\pm0.03$\\ 0.75 h
&$1.63\pm0.04$ & $0.85\pm0.05$ & $1.14\pm0.05$ & $0.86\pm0.03$ & 
$1.37\pm0.03$ &
$1.07\pm0.05$\\ 0.50 h &$1.50\pm0.04$ & $1.13\pm0.04$ & $1.14\pm0.03$ &
$0.85\pm0.03$ &$1.37\pm0.03$ & $1.19\pm0.05$\\ 0.25 h &$1.69\pm0.04$ &
$1.23\pm0.04$ & $1.31\pm0.04$ & $0.93\pm0.03$ &$1.27\pm0.04$ & $1.09\pm0.03$\\
bottom &$1.36\pm0.07$ & $0.70\pm0.03$ & $1.22\pm0.04$ & $0.68\pm0.03$ &
$1.22\pm0.03$& $0.88\pm0.02$\\ \hline \end{tabular} \end{center} \end{table}

\begin{table}[ht]

\caption{Values of the $\alpha$-exponents- from the DFA analysis of 
the shuffled
data of backscattered cross section $\eta$, Doppler velocity $<v>$ and Doppler
spectral width $\sigma_v^2$ data shown in Fig. 4(a-c); $winter$ cloud. The
correlation coefficient for all cases is around 0.997.}

\begin{center} \begin{tabular}{||c|c|c|c||} \hline position & $\eta$ &$<v>$&
$\sigma_v^2$ \\ \hline &$\alpha$ &$\alpha$ &$\alpha$ \\ \hline top
&$0.47\pm0.006$ & $0.49\pm0.006$ & $0.51\pm0.003$ \\ 0.75 h &$0.53\pm0.004$ &
$0.53\pm0.004$ & $0.54\pm0.004$ \\ 0.50 h &$0.52\pm0.004$ & $0.50\pm0.005$
&$0.52\pm0.005$ \\ 0.25 h &$0.56\pm0.006$ & $0.53\pm0.005$ & $0.48\pm0.006$ \\
bottom &$0.49\pm0.006$ & $0.50\pm0.004$ & $0.49\pm0.005$\\ \hline \end{tabular}
\end{center} \end{table}

\begin{table}[ht]

\caption{Values of the $\alpha$-exponent from the DFA analysis of shuffled data
of backscattering cross section $\eta$, Doppler velocity $<v>$ and Doppler
spectral width $\sigma_v^2$ data shown in Fig. 6(a-c); $fall$ cloud. The
correlation coefficient for all cases is around 0.994.}

\begin{center} \begin{tabular}{||c|c|c|c||} \hline position & $\eta$ & $<v>$&
$\sigma_v^2$ \\ \hline &$\alpha$ &$\alpha$ &$\alpha$ \\ \hline top
&$0.55\pm0.006$ & $0.48\pm0.006$ & $0.49\pm0.005$\\ 0.75 h &$0.48\pm0.006$ &
$0.52\pm0.007$ & $0.52\pm0.006$\\ 0.50 h &$0.50\pm0.007$ & $0.51\pm0.006$ &
$0.46\pm0.006$ \\ 0.25 h &$0.49\pm0.006$ & $0.52\pm0.005$ & $0.52\pm0.007$ \\
bottom &$0.51\pm0.010$ & $0.51\pm0.005$ & $0.53\pm0.007$\\ \hline \end{tabular}
\end{center} \end{table}

\newpage
\newpage \begin{figure}[ht] \begin{center} \leavevmode 
\epsfysize=4.5cm
\epsffile{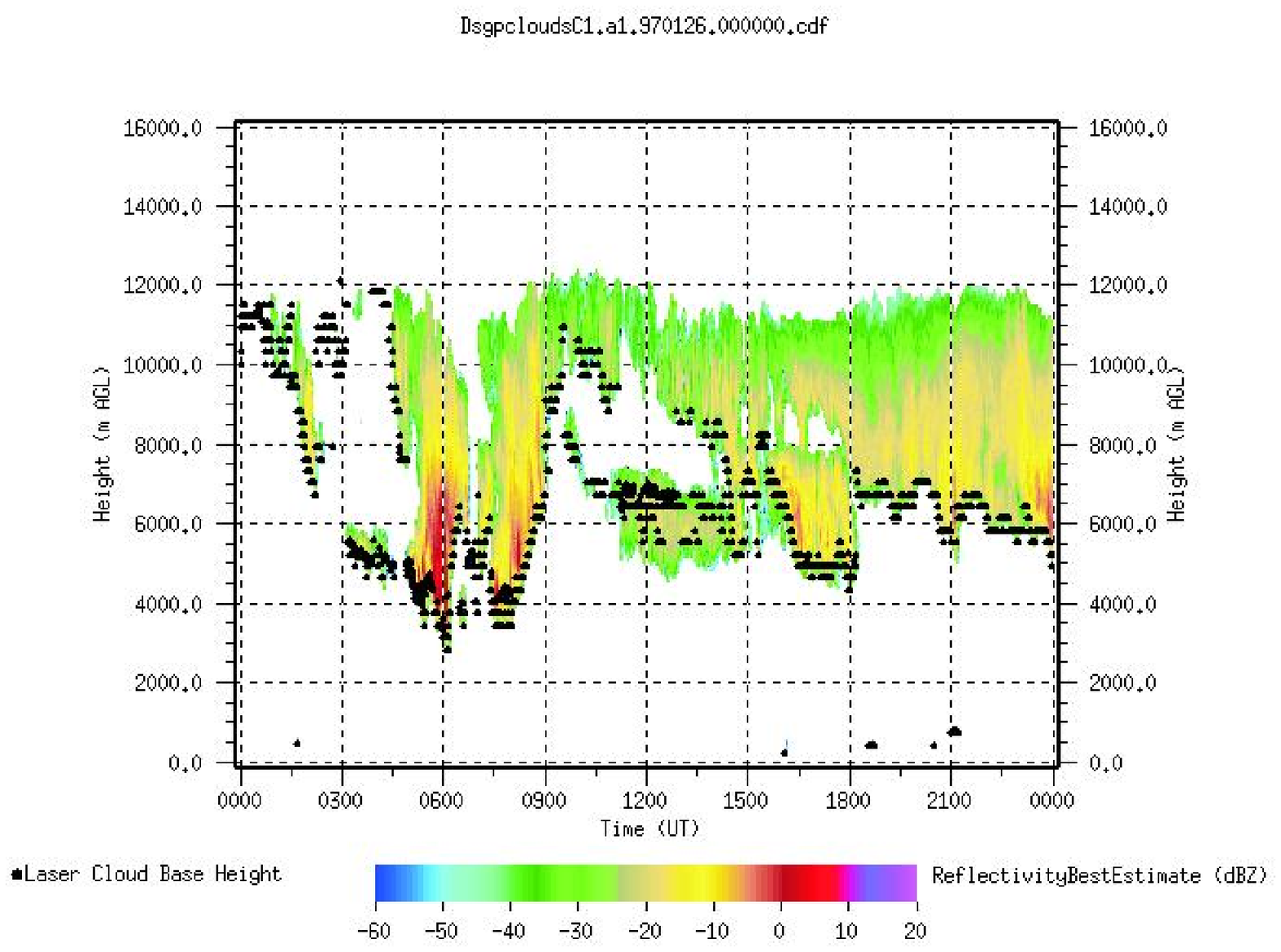} 
\hfill 
\leavevmode \epsfysize=4.5cm 
\epsffile{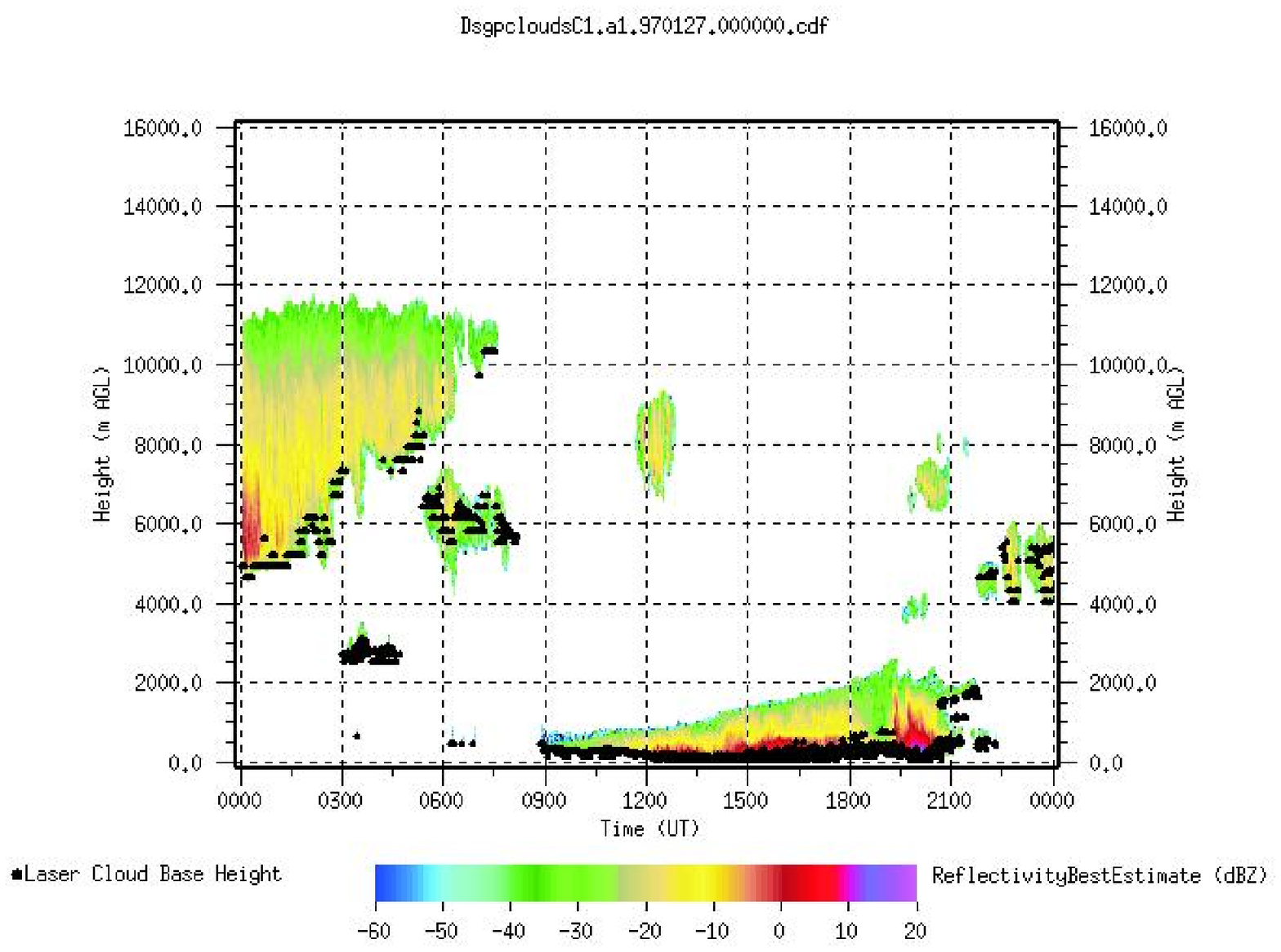}
\vfill \leavevmode \epsfysize=4.5cm 
\epsffile{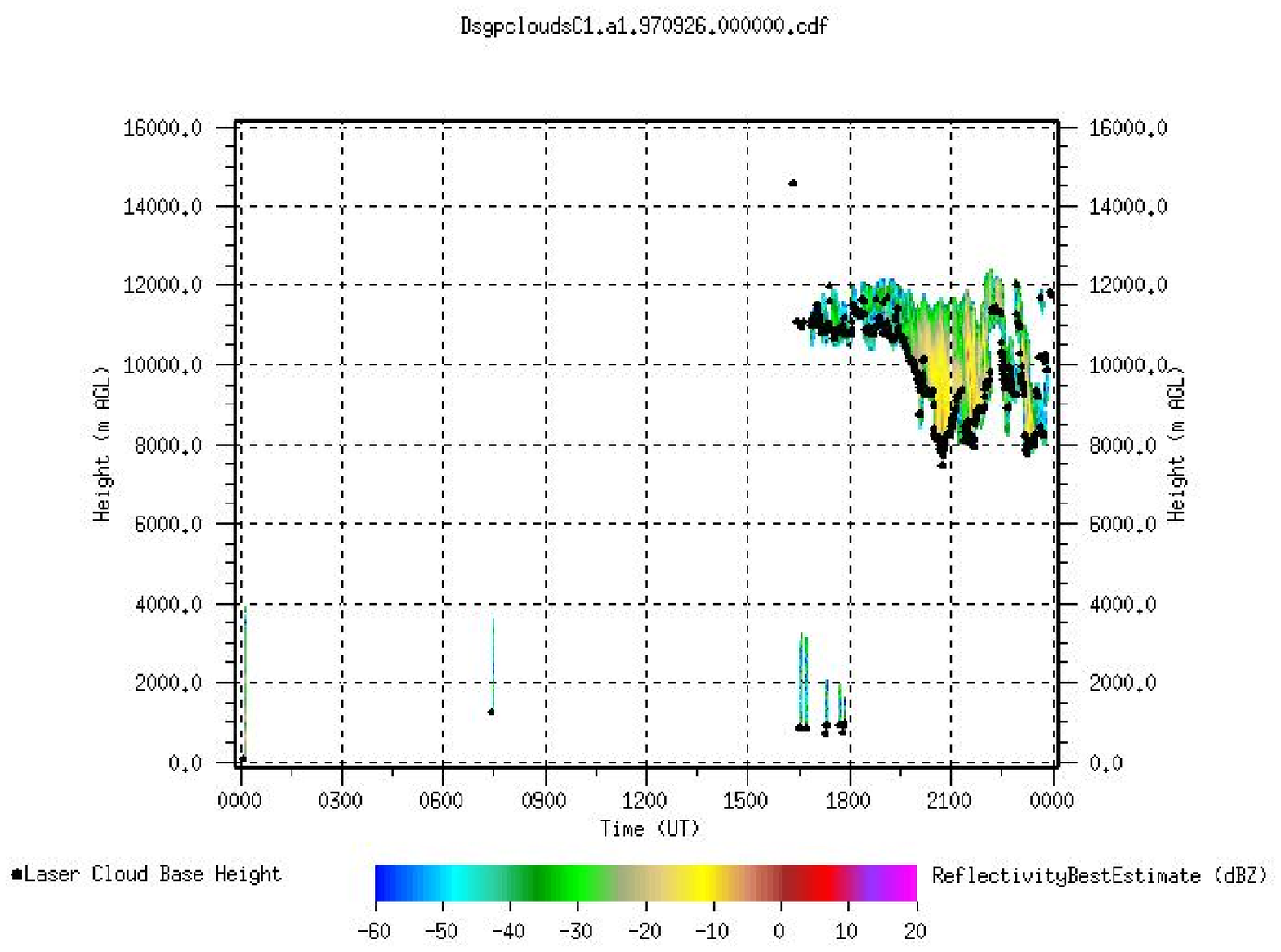} 
\end{center}
\label{fig1} \caption{Radar reflectivity observations on (a) Jan. 26, (b) 27,
1997 and (c) on Sept. 26, 1997 at the Southern Great Plains site of Atmospheric
Radiation Measurements program.} \end{figure}

\newpage \begin{figure}[ht] \begin{center} \leavevmode \epsfysize=8cm
\epsffile{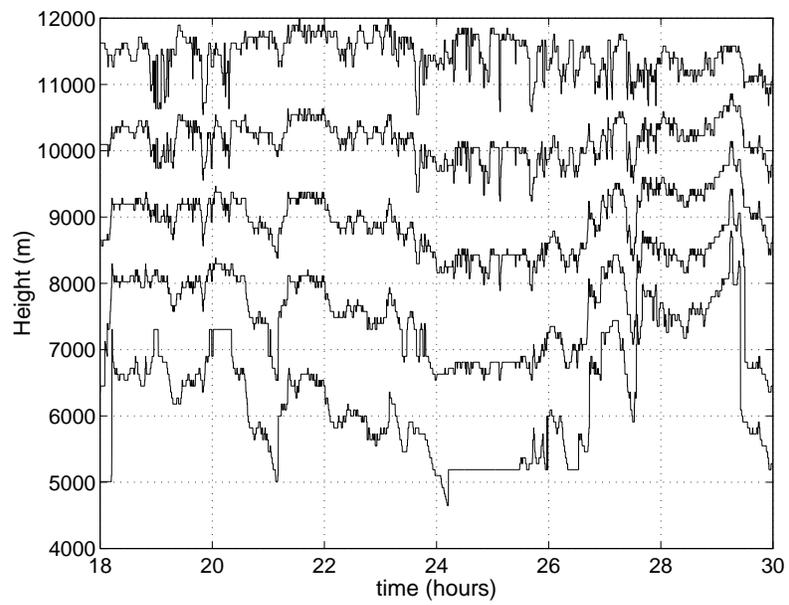} \end{center} \label{fig2} \caption{Cloud 
profiles at top and
bottom and at $0.75$, $0.5$ and $0.25$ of the thickness $h$ of the $winter$
cirrus cloud (Fig.1a,b) along which the backscattering cross section, Doppler
velocity and Doppler spectral width are studied.} \end{figure}

\newpage \begin{figure}[ht] \begin{center} \leavevmode \epsfysize=8cm
\epsffile{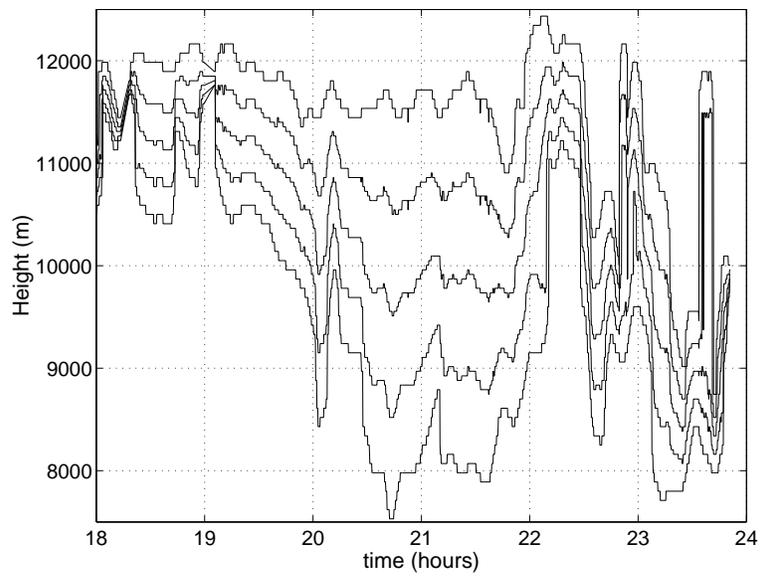} \end{center} \label{fig3} \caption{Cloud 
profiles at top and
bottom and at $0.75$, $0.5$ and $0.25$ of the thickness $h$ of the 
$fall$ cirrus
cloud along which the backscattering cross section, Doppler velocity 
and Doppler
spectral width are studied. Measurements are obtained at the Southern Great
Plains site of Atmospheric Radiation Measurements program on Sept. 
26, 1997 (Fig.
1c).} \end{figure}

\newpage \begin{figure}[ht] \begin{center} \leavevmode \epsfysize=8cm
\epsffile{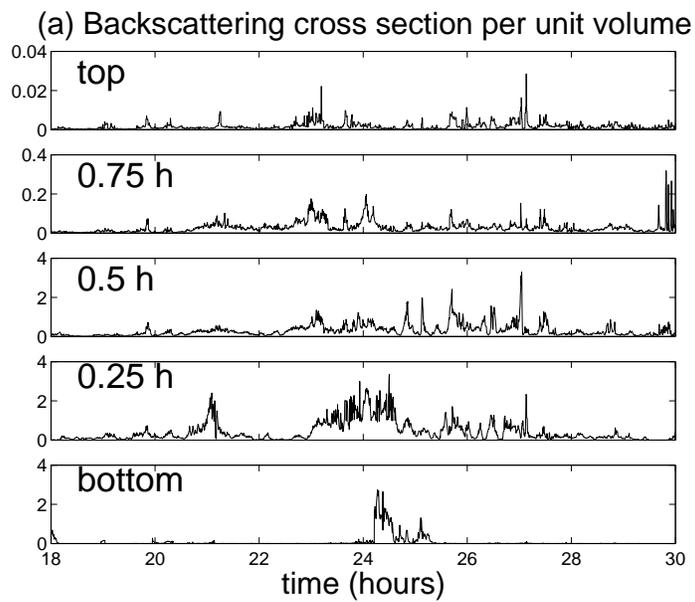} \end{center} \label{fig4a} \caption{(a) Backscattering
cross section per unit volume at different relative heights to the 
thickness $h$
of the cirrus cloud as measured on Jan. 26 and 27, 1997 at the Southern Great
Plains site of Atmospheric Radiation Measurements program (data in Fig. 1a,b).}
\end{figure}

\newpage \begin{figure}[ht] \begin{center} \leavevmode \epsfysize=8cm
\epsffile{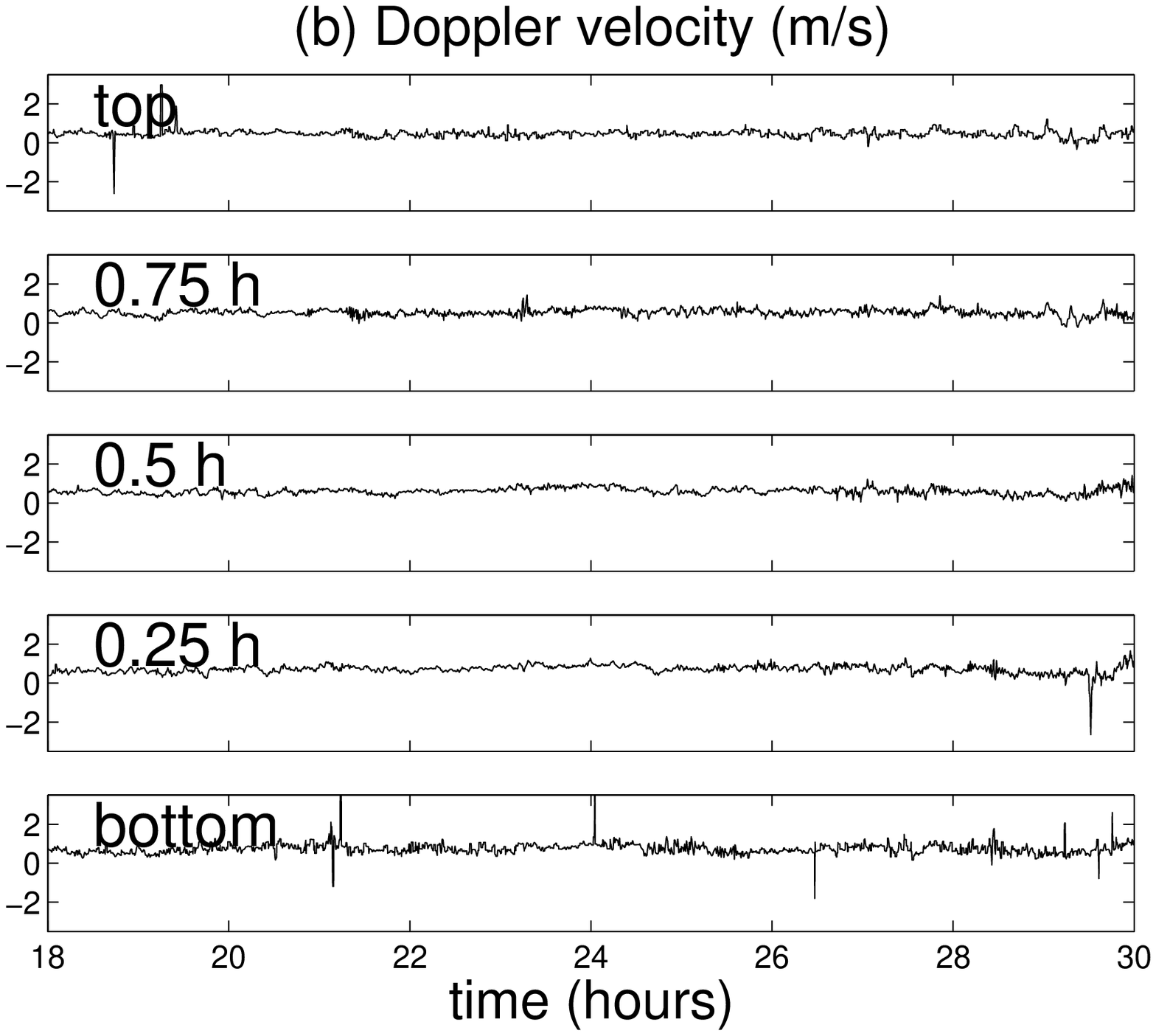} \end{center} \label{fig4b}
\nopagebreak
Figure
4: (b) Doppler velocity (m/s). The positive spike between 20 and 22 h of the
"bottom" data series is of order of 11~m/s (not shown) at heights in the cloud
shown in Fig. 2
\end{figure} 
 
\newpage \begin{figure}[ht] \begin{center} \leavevmode \epsfysize=8cm
\epsffile{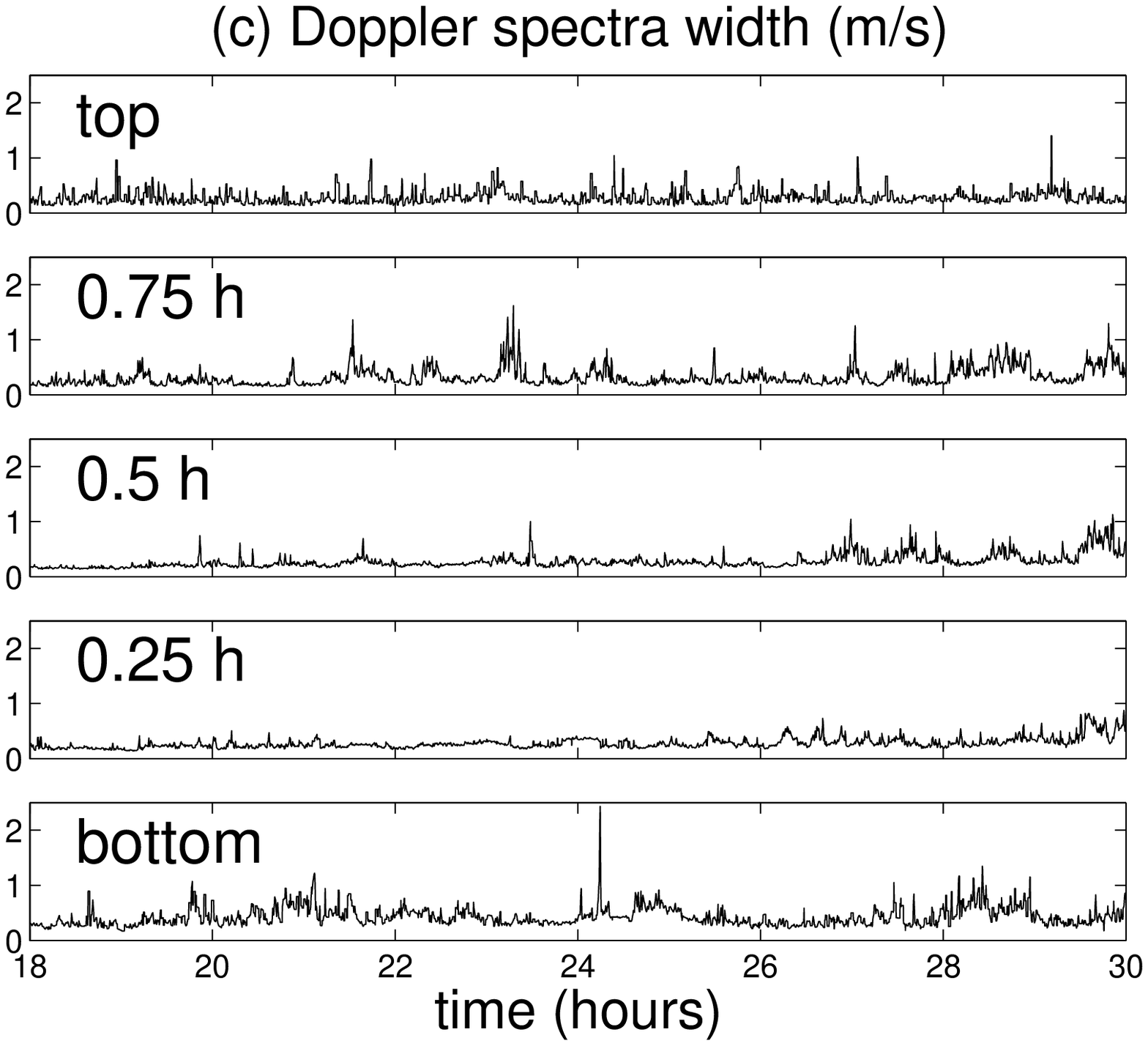} \end{center} \label{fig4c} 
\nopagebreak Figure
4: (c) Doppler spectral width (m/s) at heights in the cloud shown in Fig. 2
\end{figure} 

\newpage \begin{figure}[ht] \begin{center} \leavevmode \epsfysize=4.5cm
\epsffile{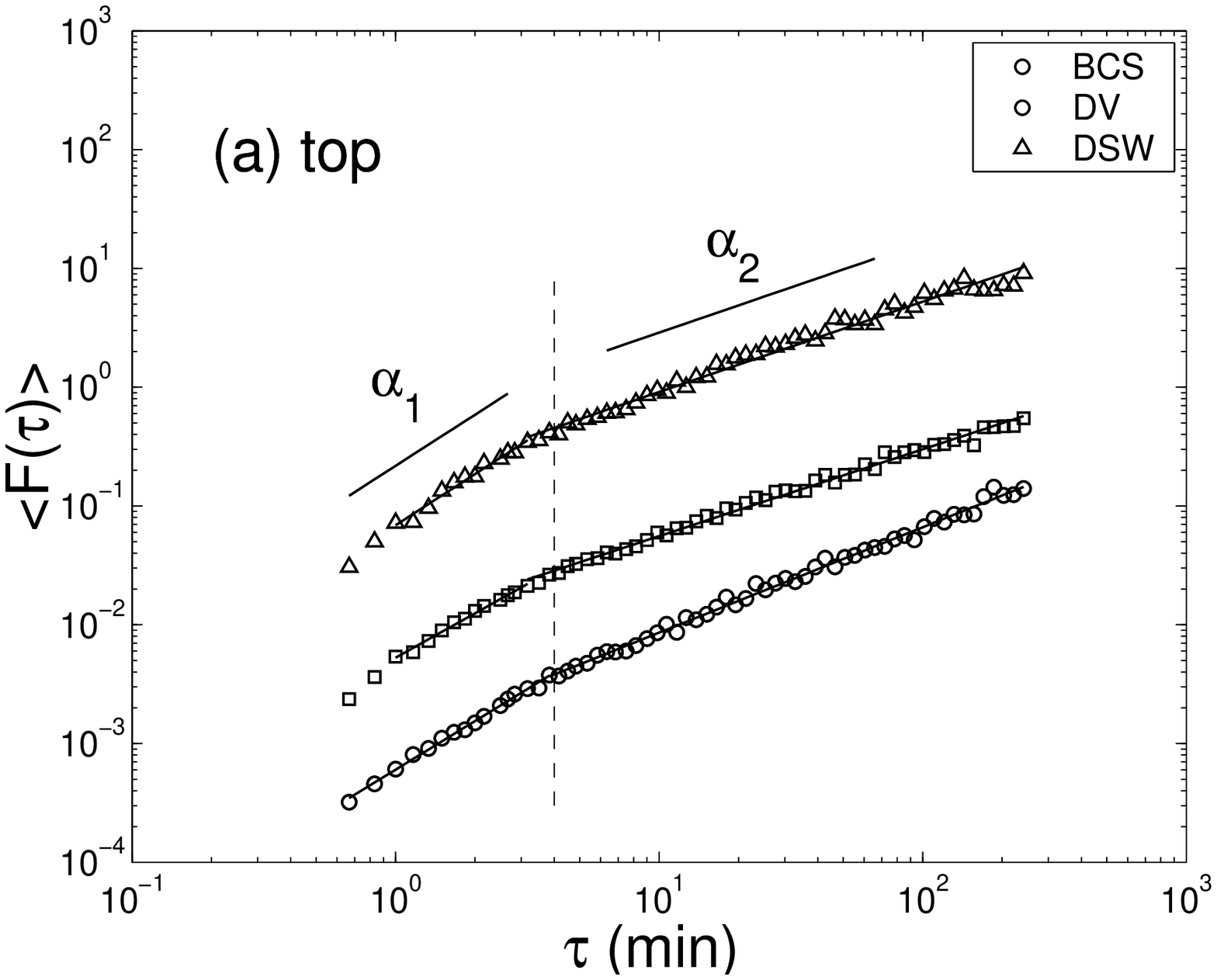} \hfill \leavevmode \epsfysize=4.5cm
\epsffile{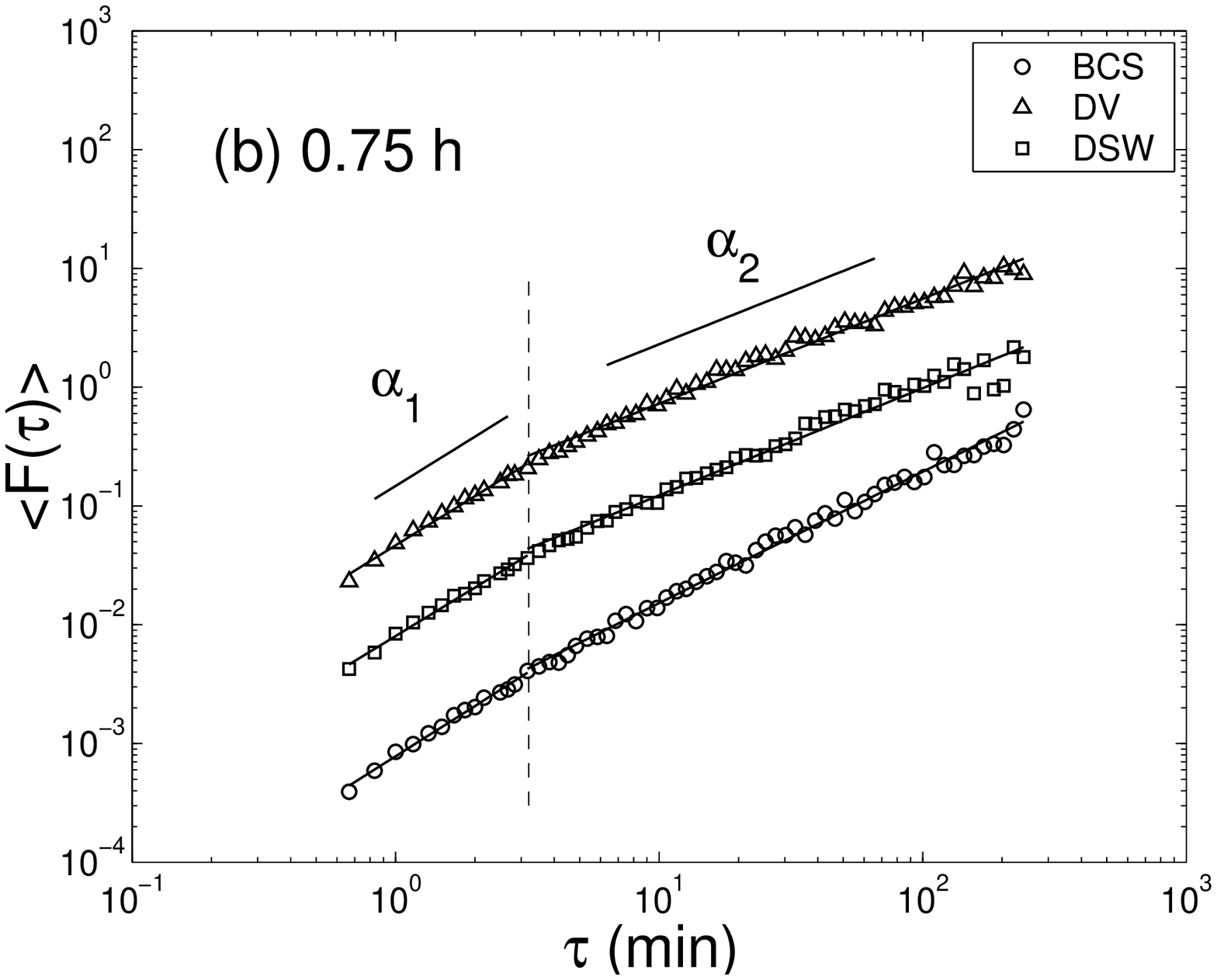} \vfill \leavevmode \epsfysize=4.5cm
\epsffile{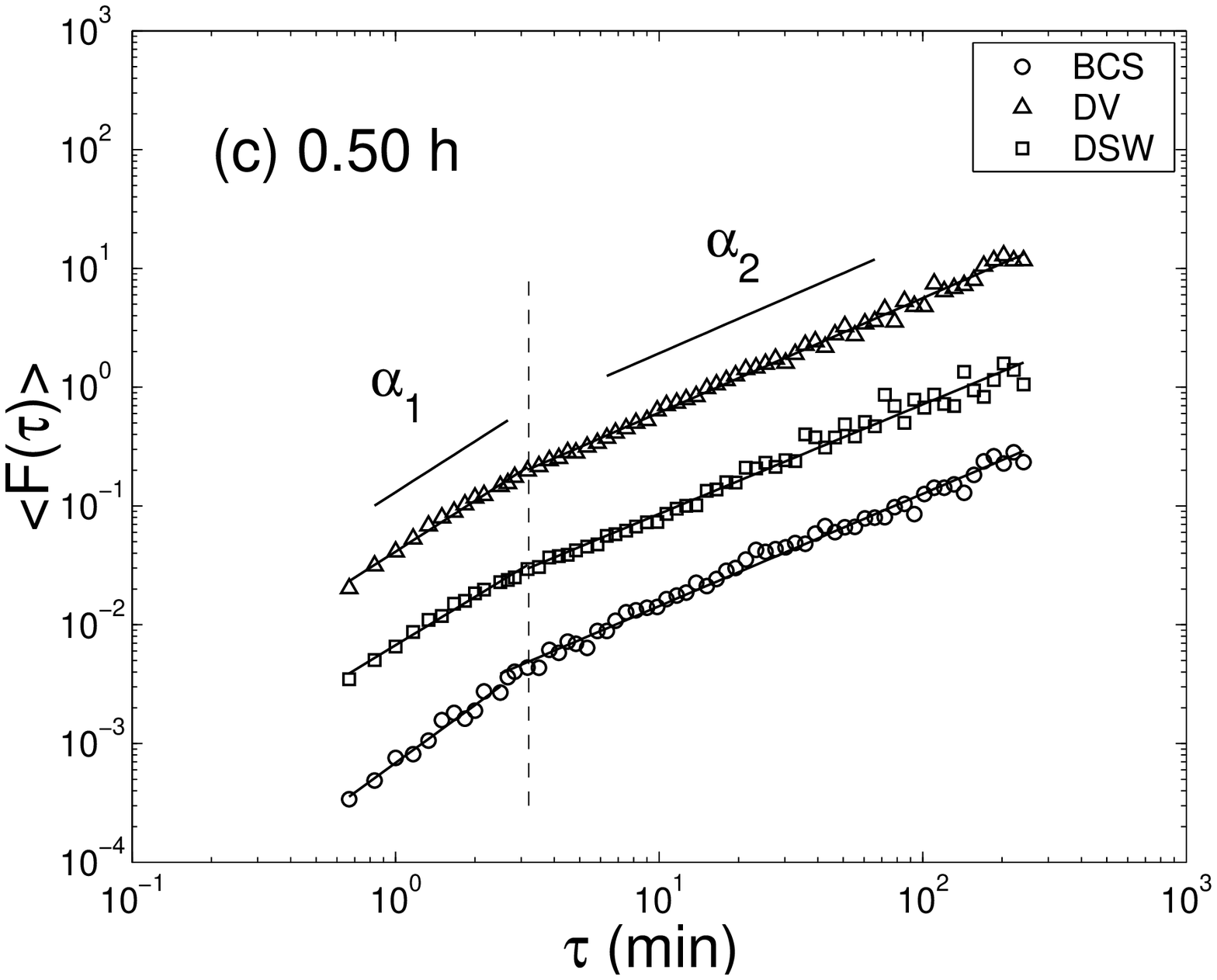} \vfill \leavevmode \epsfysize=4.5cm
\epsffile{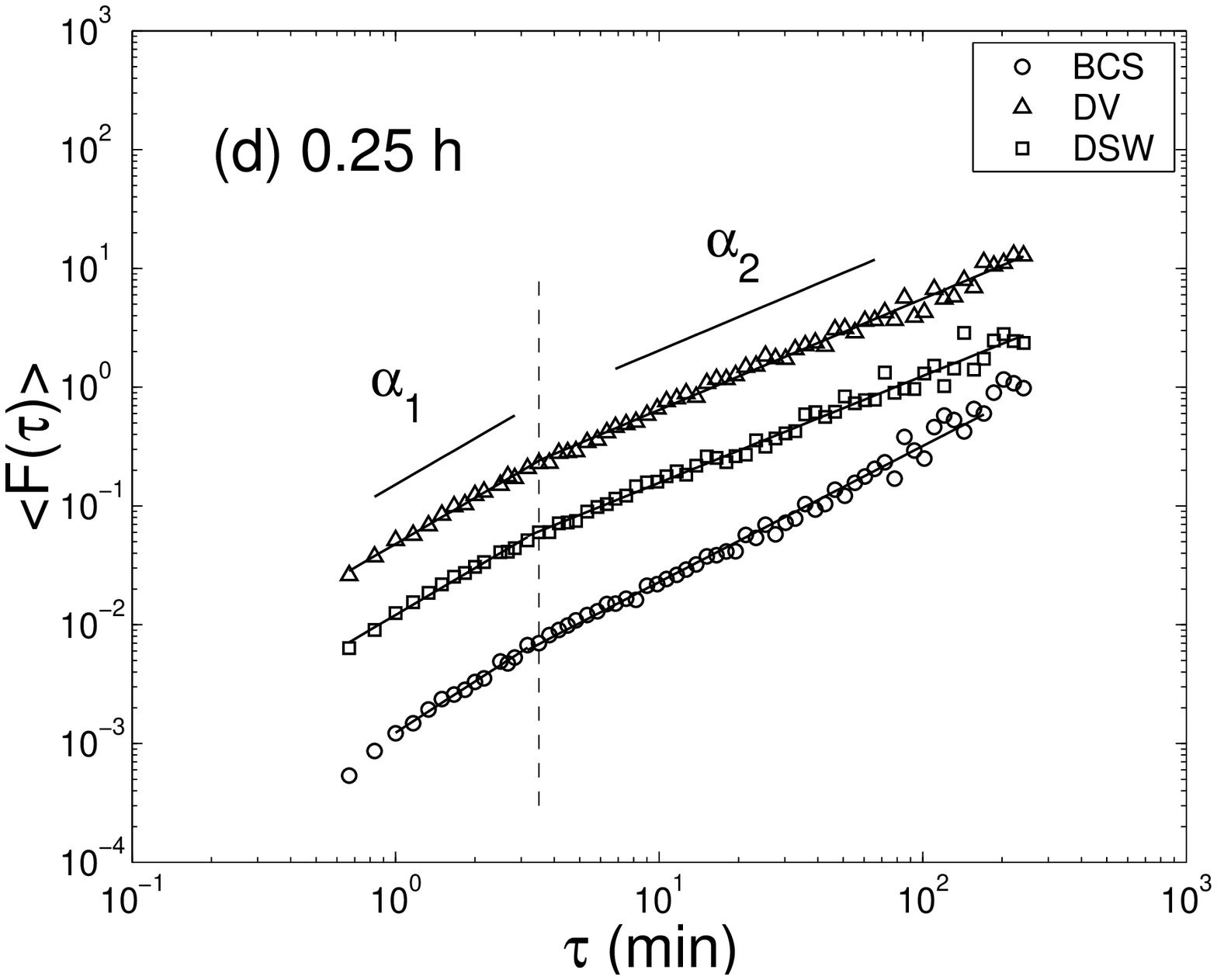} \hfill \leavevmode \epsfysize=4.5cm
\epsffile{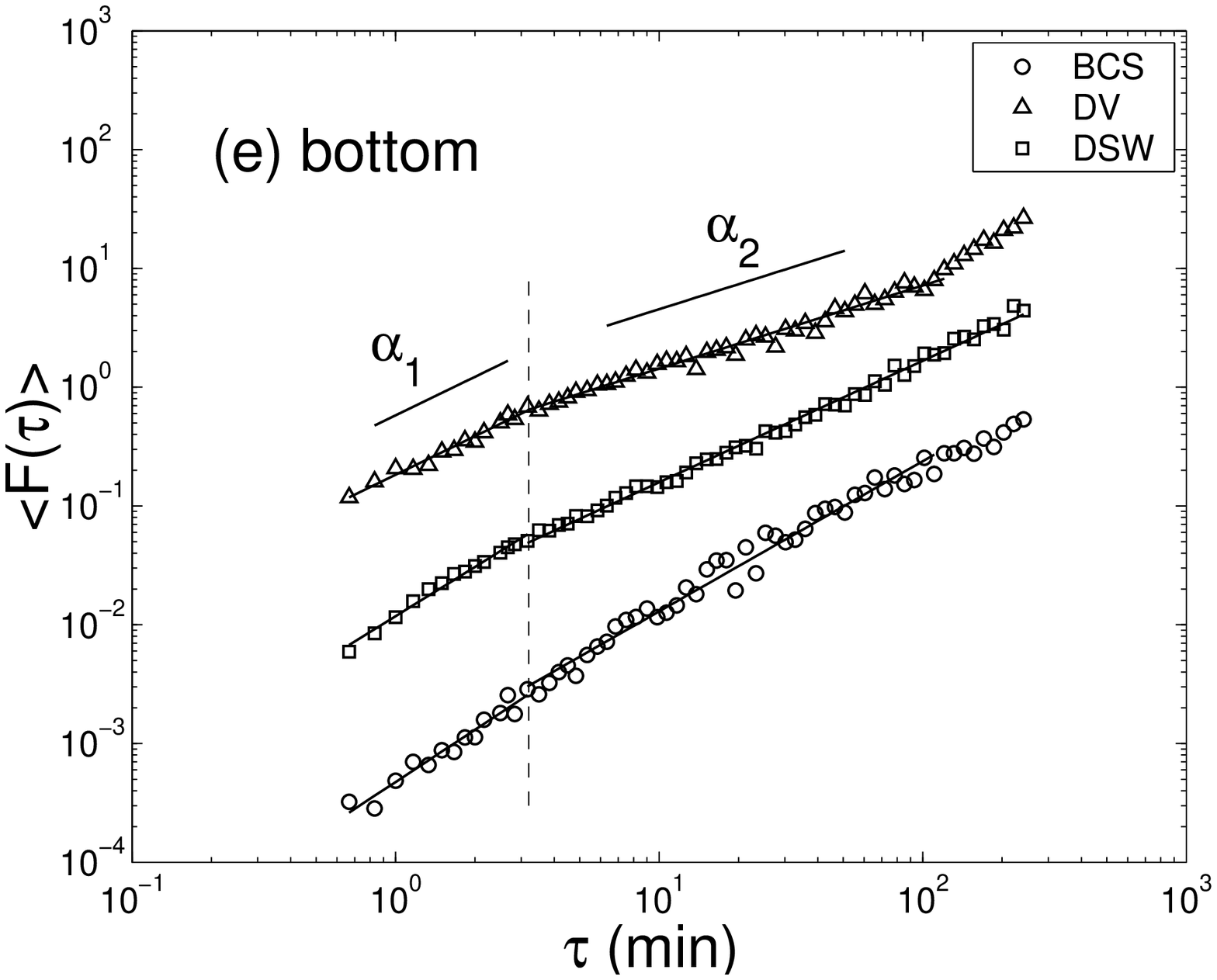} \end{center} \label{fig5} \caption{DFA-functions for
backscattering cross section (circles), Doppler velocity (triangles) 
and Doppler
spectral width (squares) at the (a) top, (b) 0.75, (c) 0.50, (d) 0.25 
relative to
the thickness $h$ of the cloud, and (e) bottom of the $winter$ cloud shown in
Fig. 2. DFA-functions are displaced for readability. The $\alpha$-values are
listed in Table 1.} \end{figure}

\newpage \begin{figure}[ht] \begin{center} \leavevmode \epsfysize=8cm
\epsffile{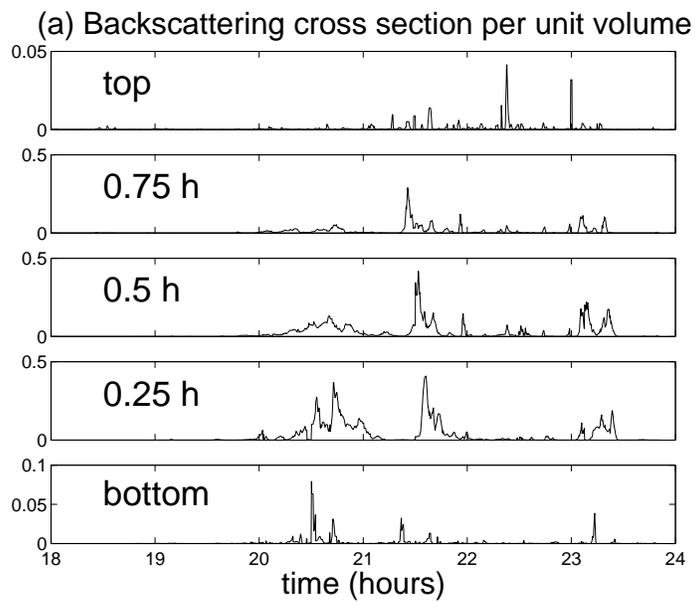} 
\end{center} \label{fig6} \caption{(a) 
Backscattering cross
section per unit volume at different relative heights to the 
thickness $h$ of the
cirrus cloud as measured on Sept. 26, 1997 at the Southern Great Plains site of
Atmospheric Radiation Measurements program (Fig. 1c).} \end{figure}

\newpage \begin{figure}[ht] \begin{center} \leavevmode \epsfysize=8cm
\epsffile{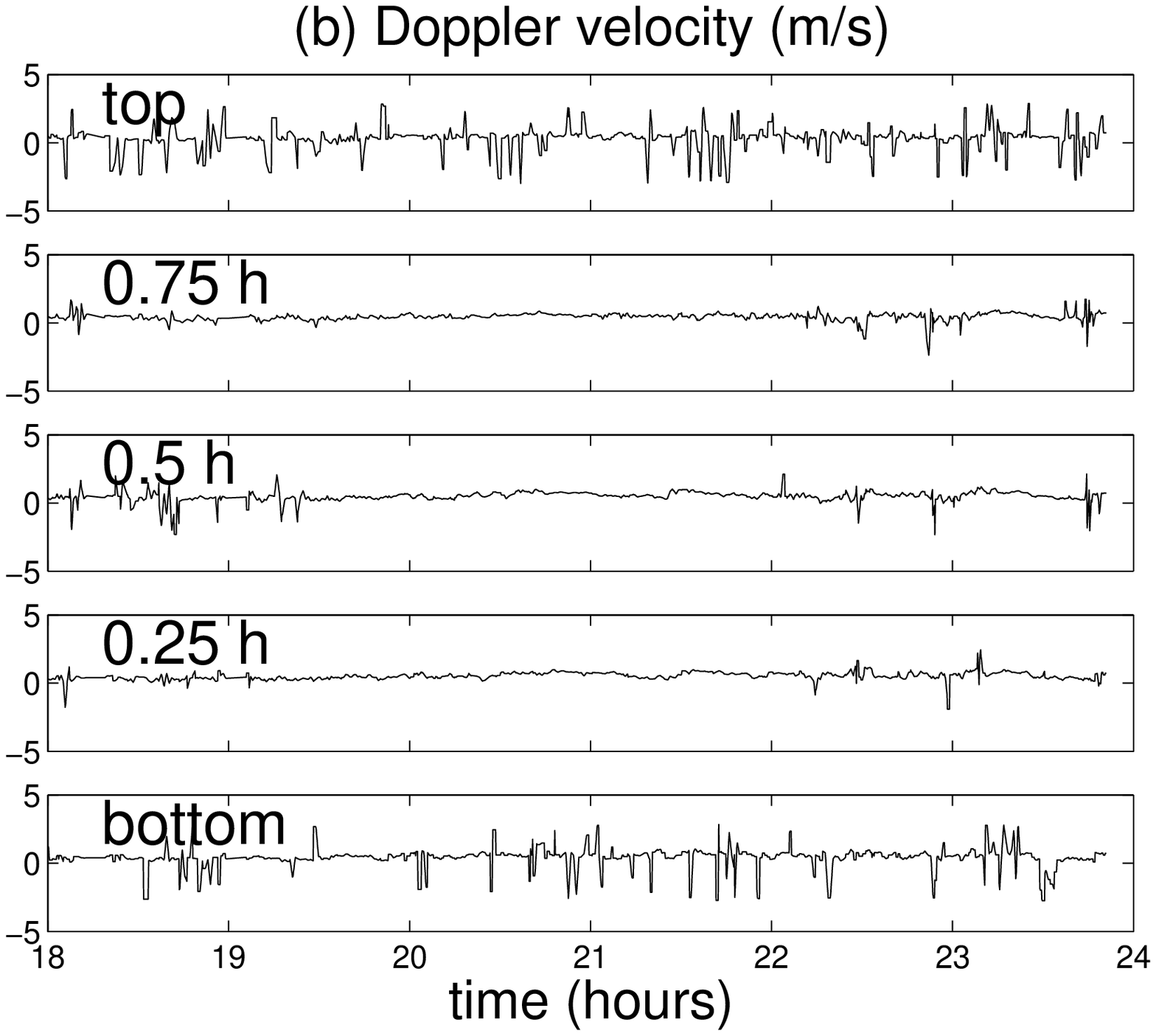}
\end{center} \nopagebreak Figure 6: (b) Doppler velocity
(m/s) at heights in the cloud shown in Fig. 3. \end{figure}

\newpage \begin{figure}[ht] \begin{center} \leavevmode \epsfysize=8cm
\epsffile{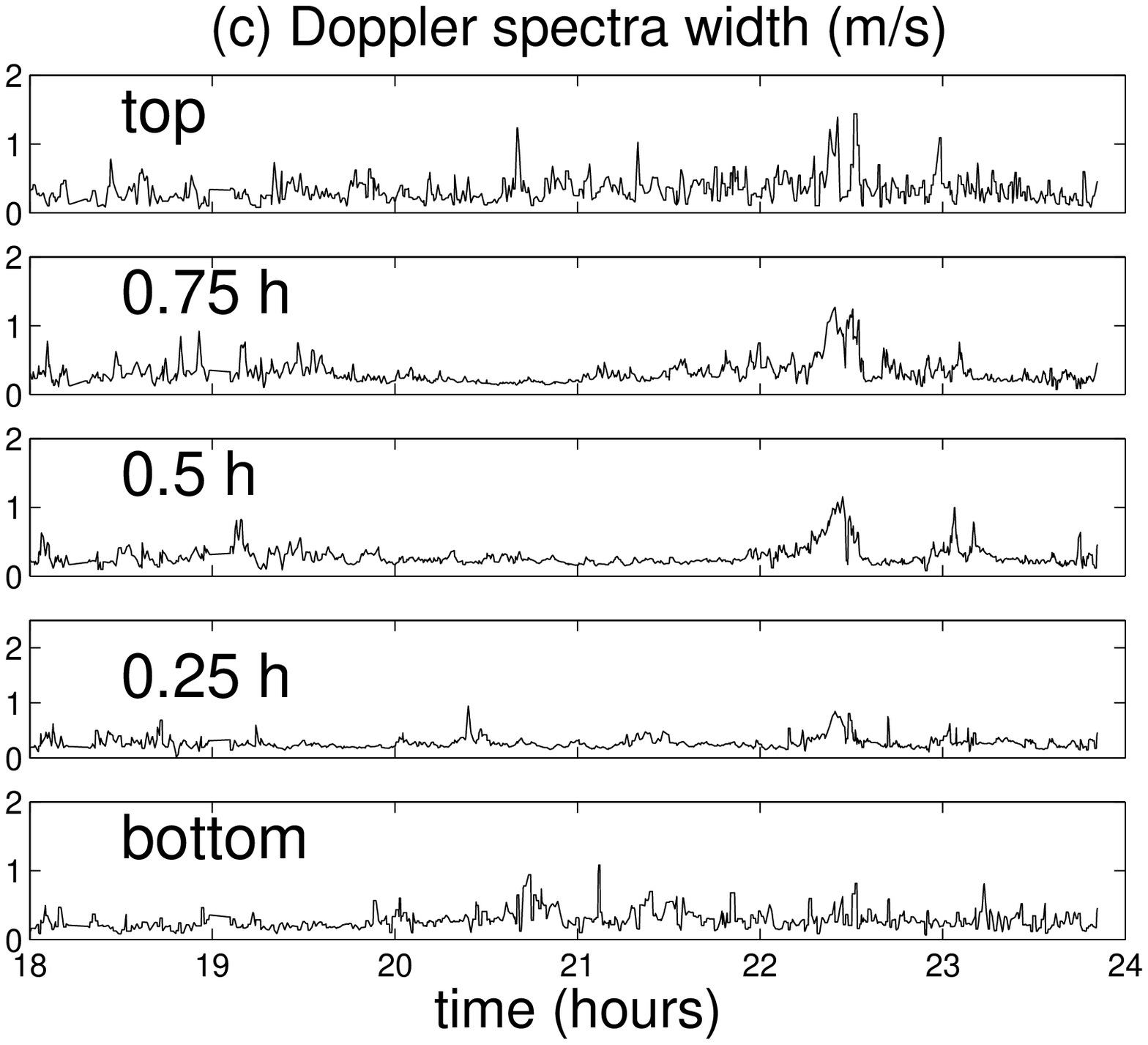} 
\end{center} \nopagebreak Figure 6: (c) Doppler spectral
width (m/s) at heights in the cloud shown in Fig. 3. \end{figure}

\newpage \begin{figure}[ht] \begin{center} \leavevmode \epsfysize=4.5cm
\epsffile{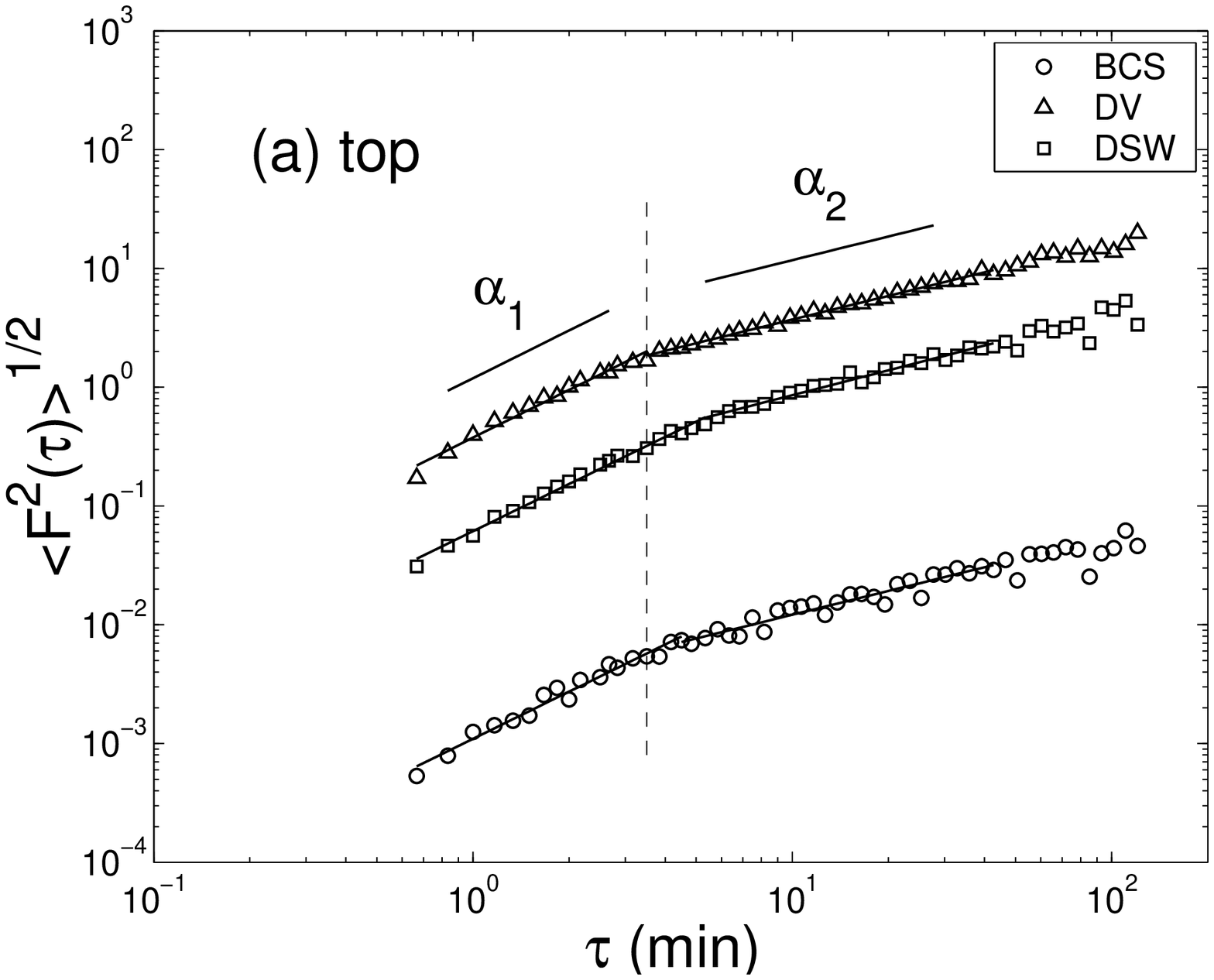} \hfill \leavevmode \epsfysize=4.5cm 
\epsffile{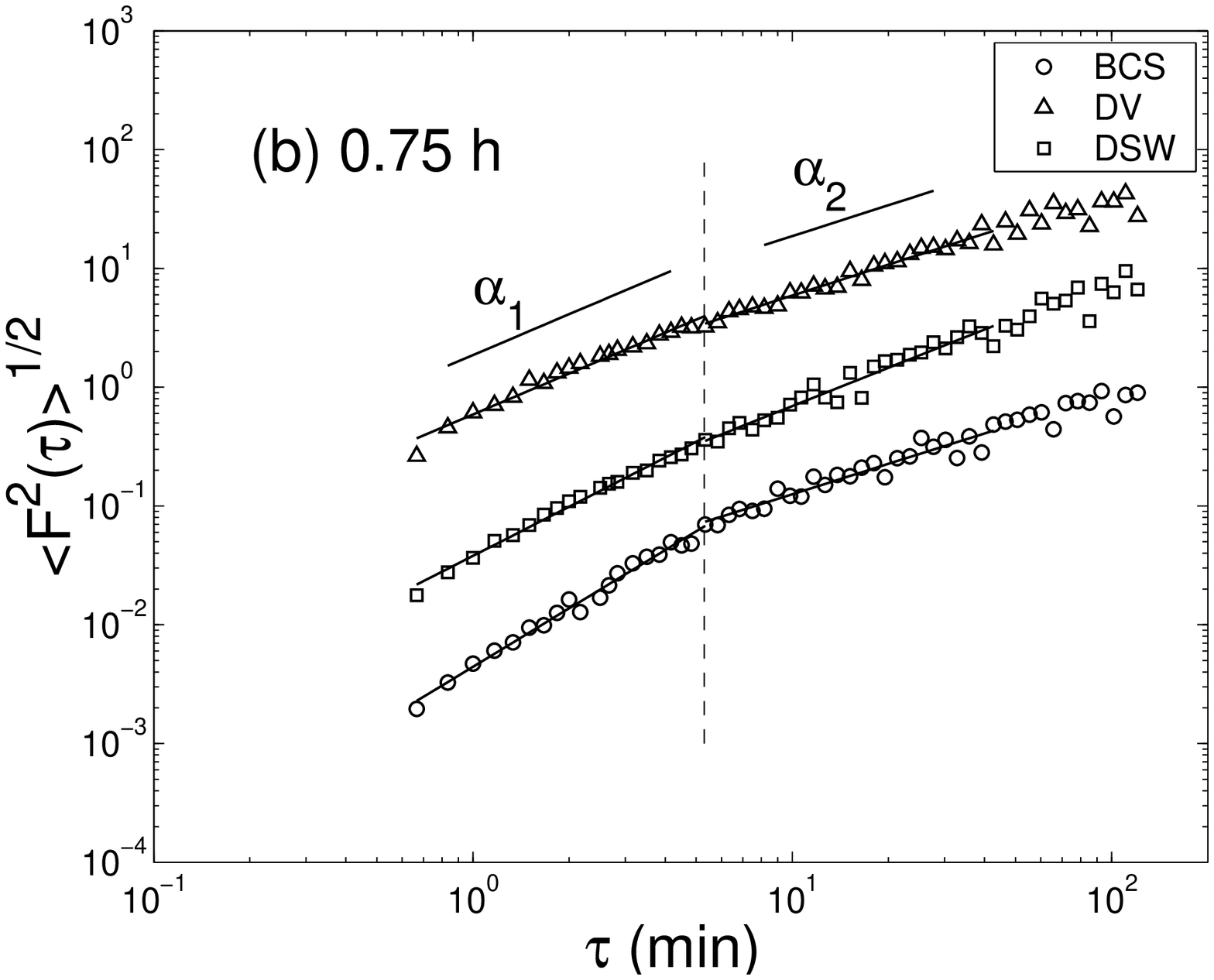}
\vfill \leavevmode \epsfysize=4.5cm \epsffile{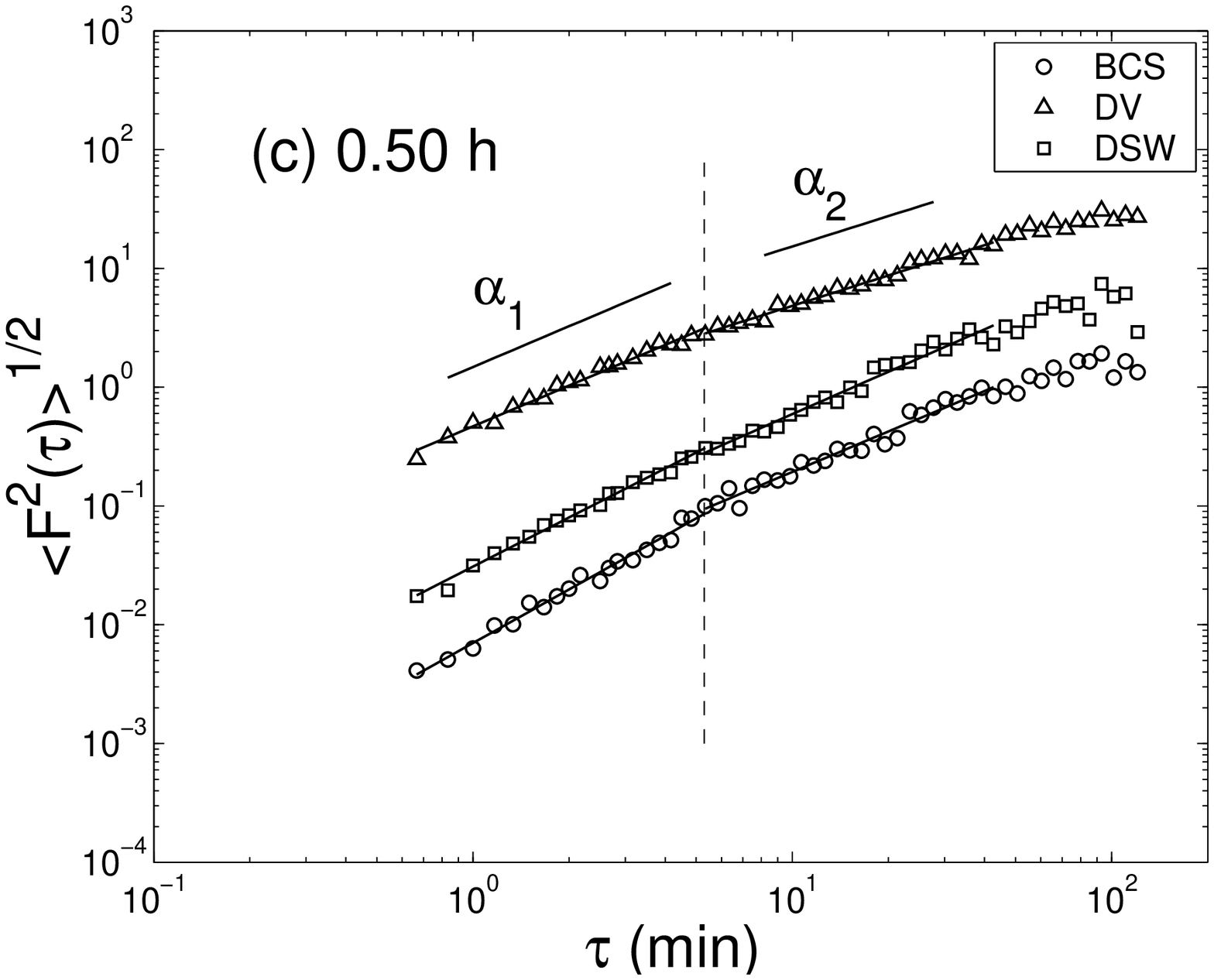} \vfill \leavevmode
\epsfysize=4.5cm \epsffile{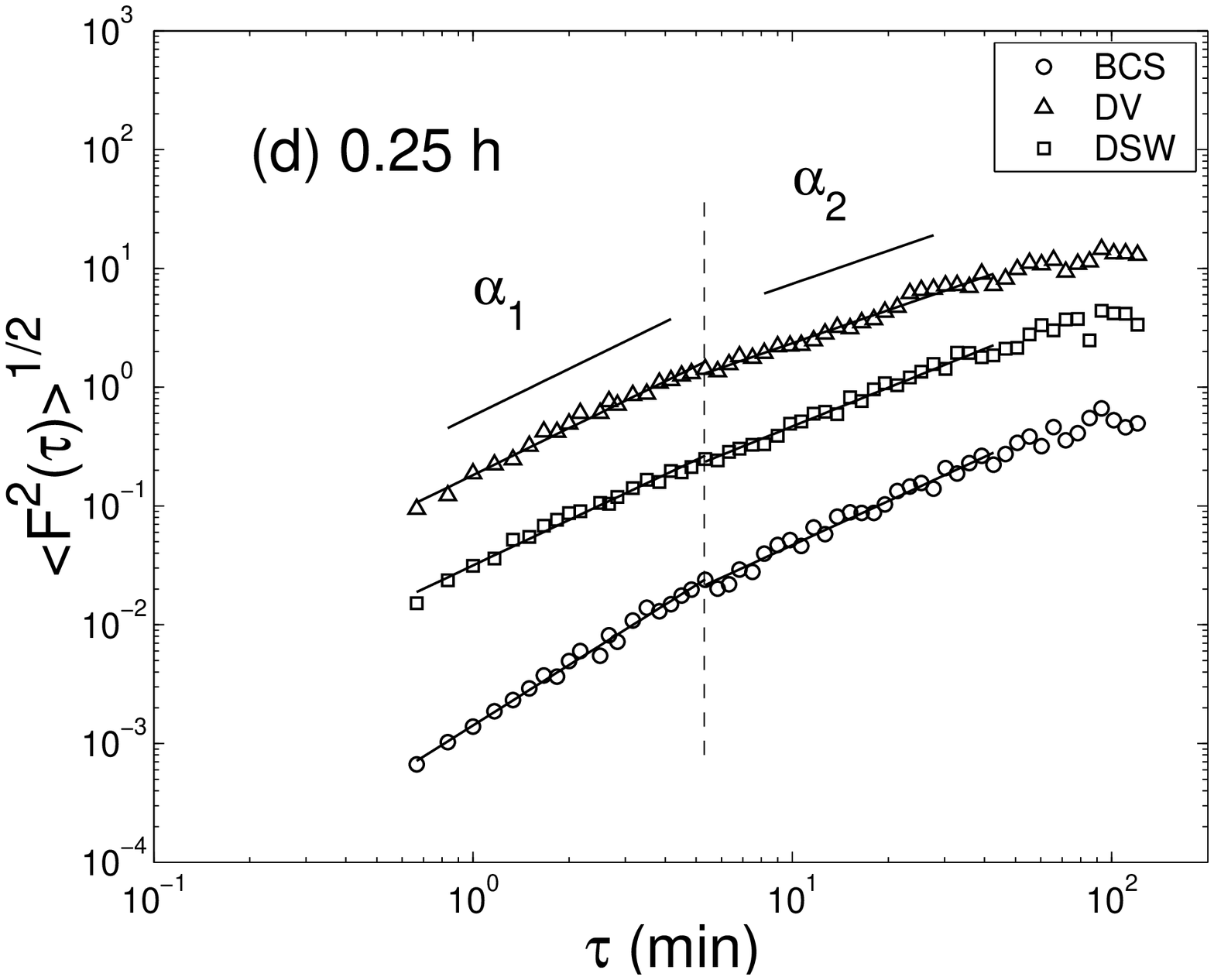} \hfill \leavevmode \epsfysize=4.5cm
\epsffile{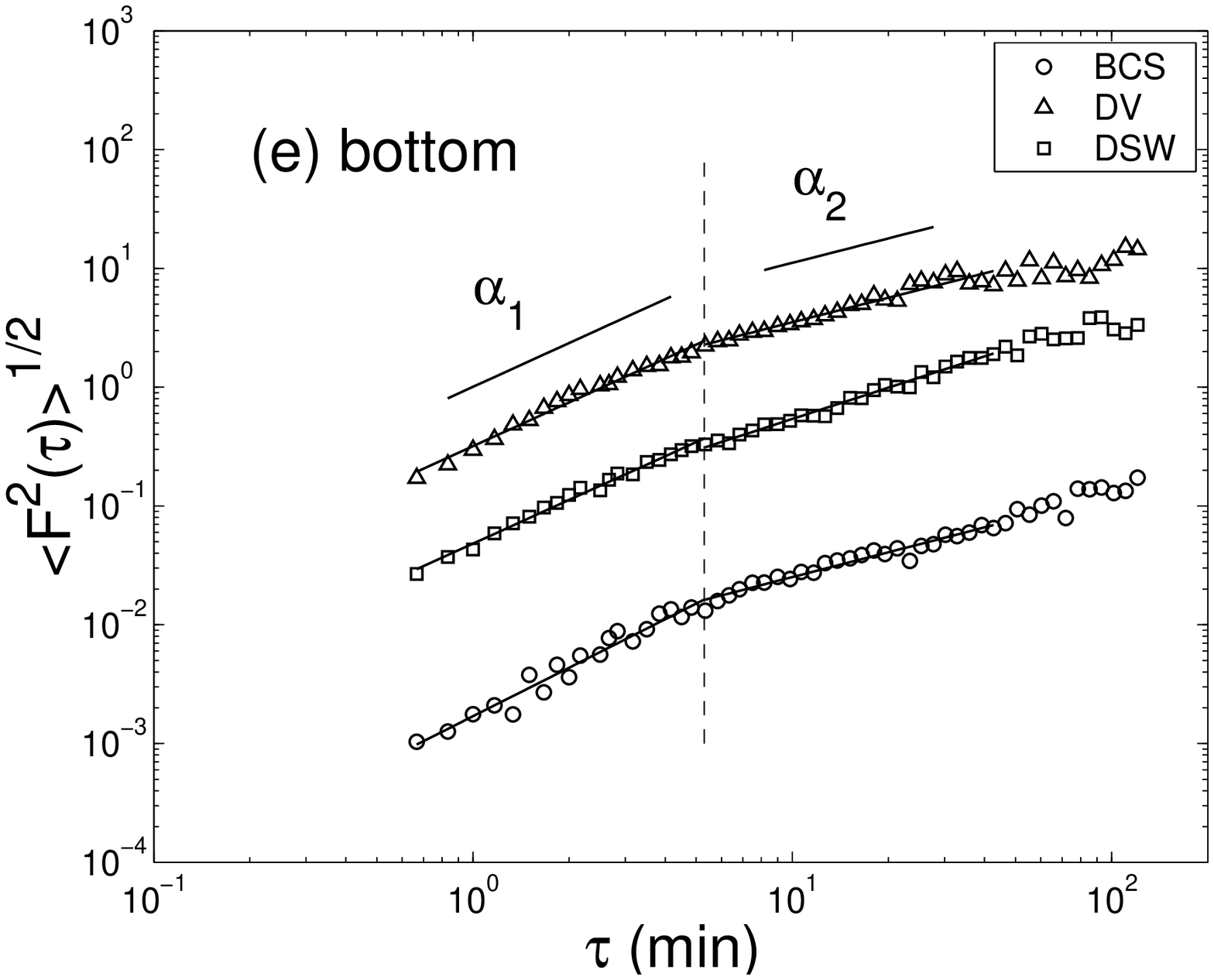} \end{center} \label{fig7} \caption{DFA-functions for
backscattering cross section (circles), Doppler velocity (triangles) 
and Doppler
spectral width (squares) at the (a) top, (b) 0.75, (c) 0.50, (d) 0.25 
relative to
the thickness $h$ of the cloud, and (e) bottom of the cloud shown in Fig. 3.
DFA-functions are displaced for readability. The $\alpha$-values are listed in
Table 2.} \end{figure}

\newpage \begin{figure}[ht] \begin{center} \leavevmode \epsfysize=4.5cm
\epsffile{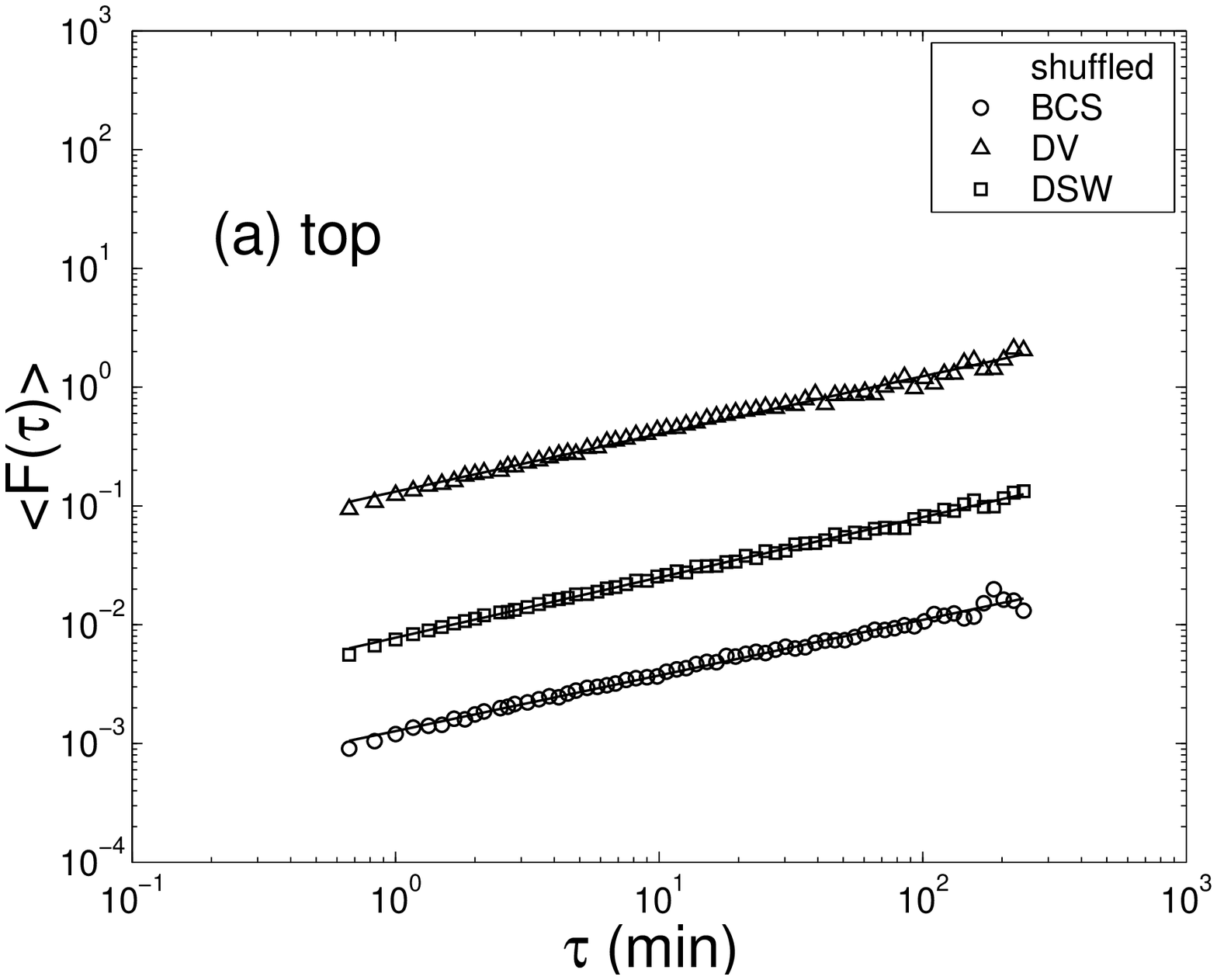} \hfill \leavevmode \epsfysize=4.5cm \epsffile{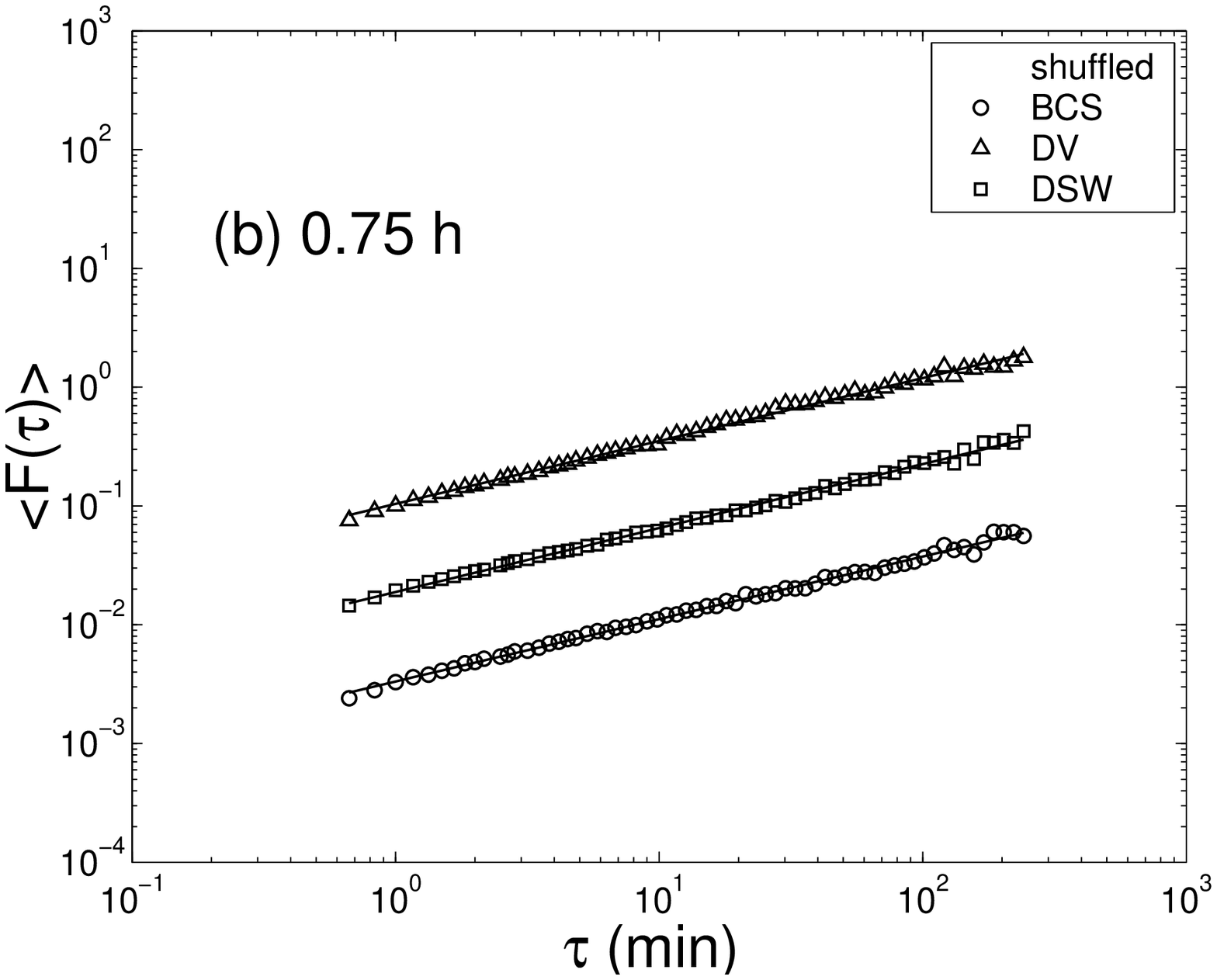}
\vfill \leavevmode \epsfysize=4.5cm \epsffile{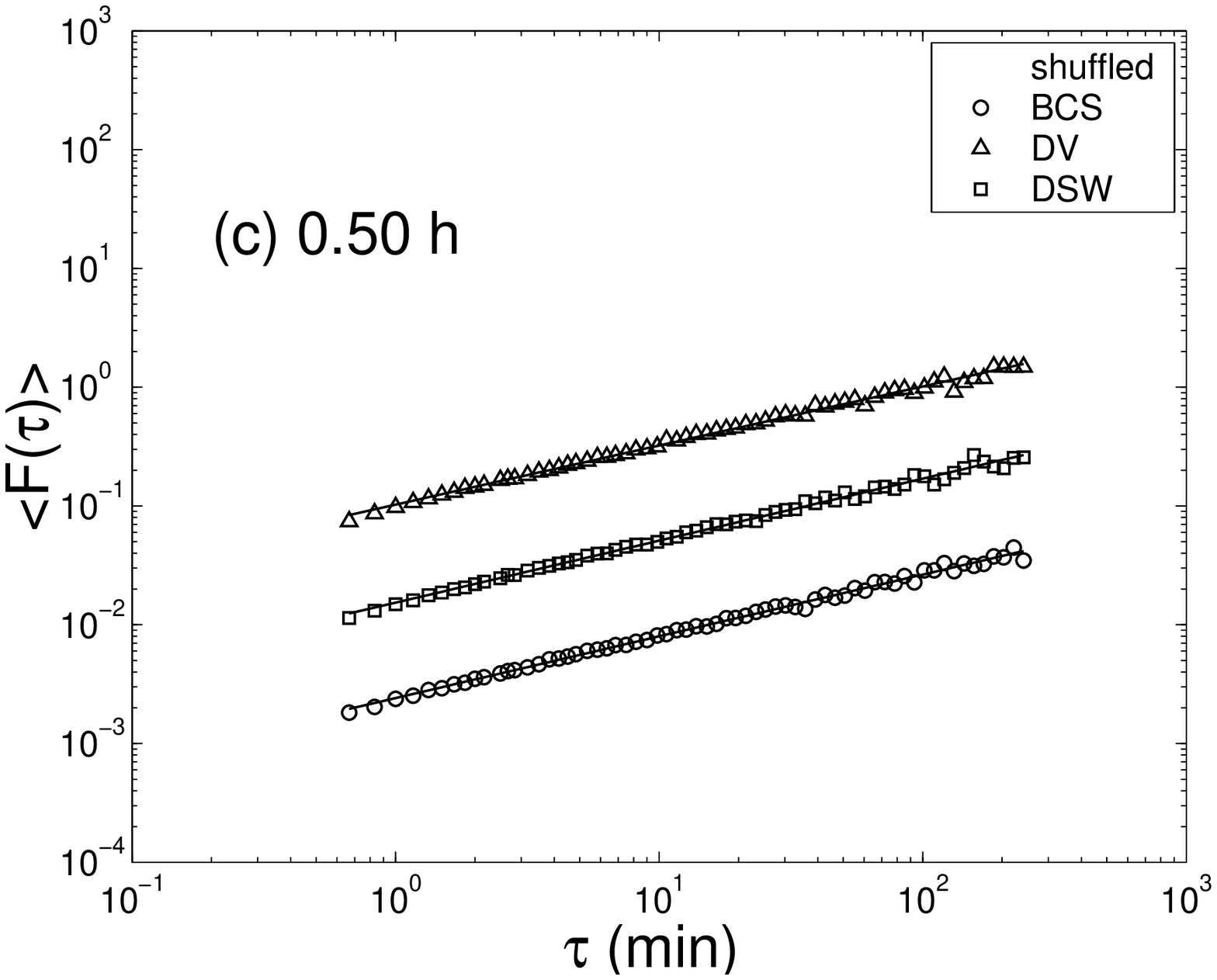} \vfill \leavevmode
\epsfysize=4.5cm \epsffile{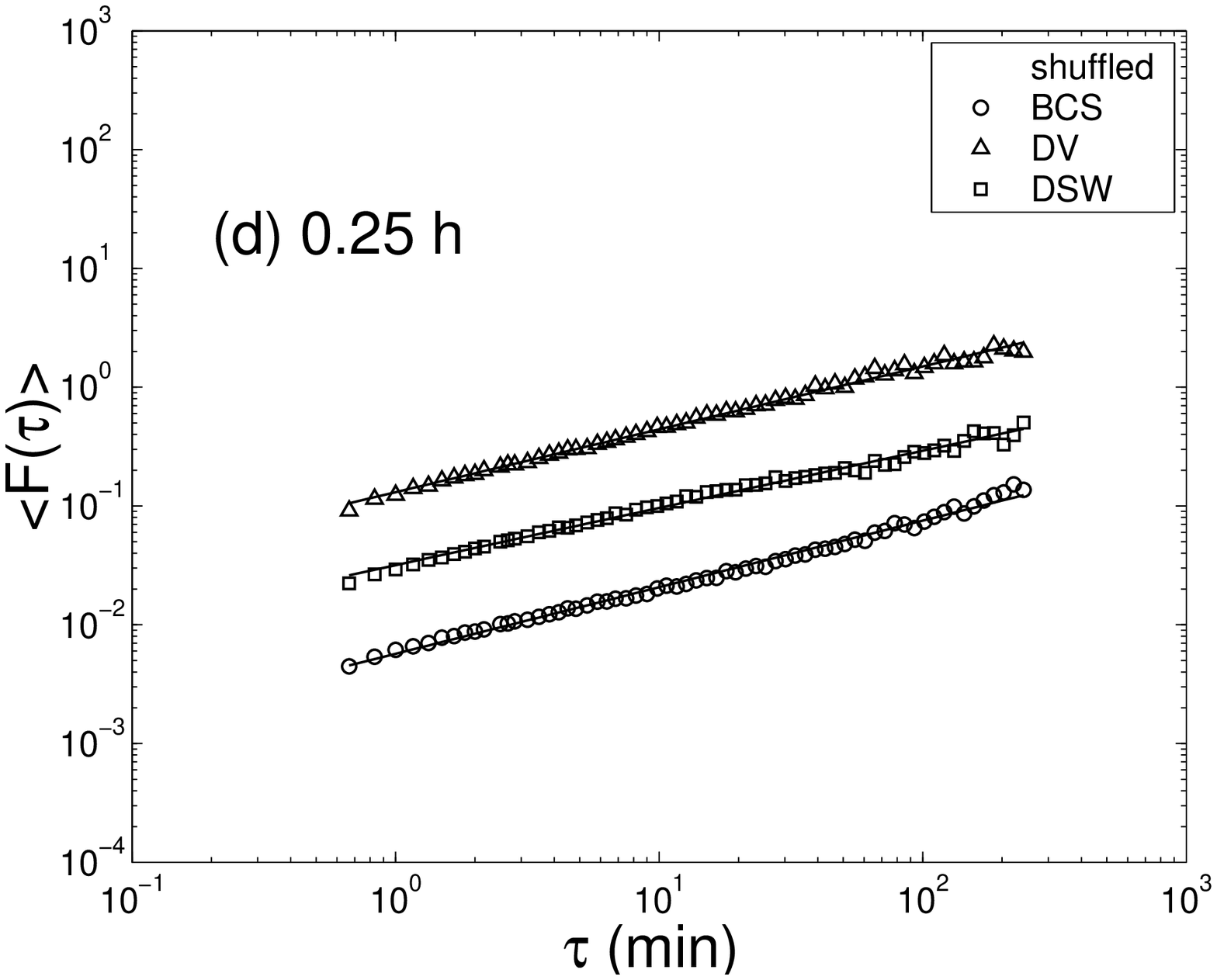} \hfill \leavevmode \epsfysize=4.5cm
\epsffile{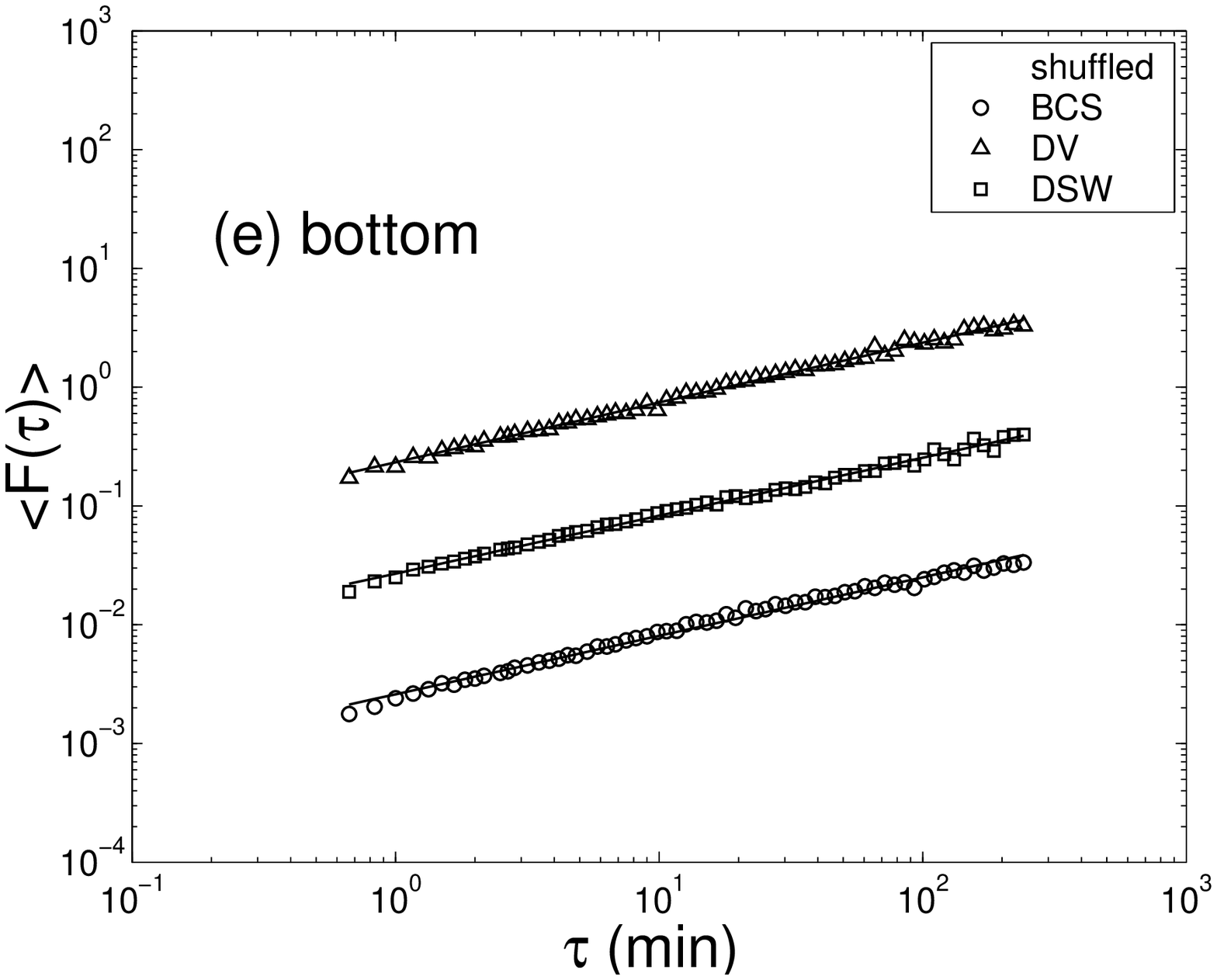} \end{center} \label{fig8} \caption{DFA-functions for
shuffled data of: backscattering cross section (circles), Doppler velocity
(triangles) and Doppler spectral width (squares) at the (a) top, (b) 0.75, (c)
0.50, (d) 0.25 relative to the thickness $h$ of the cloud, and (e) 
bottom of the
$winter$ cloud shown in Fig. 2. DFA-functions are displaced for 
readability. The
$\alpha$-values are listed in Table 3.} \end{figure}

\newpage \begin{figure}[ht] \begin{center} \leavevmode \epsfysize=4.5cm
\epsffile{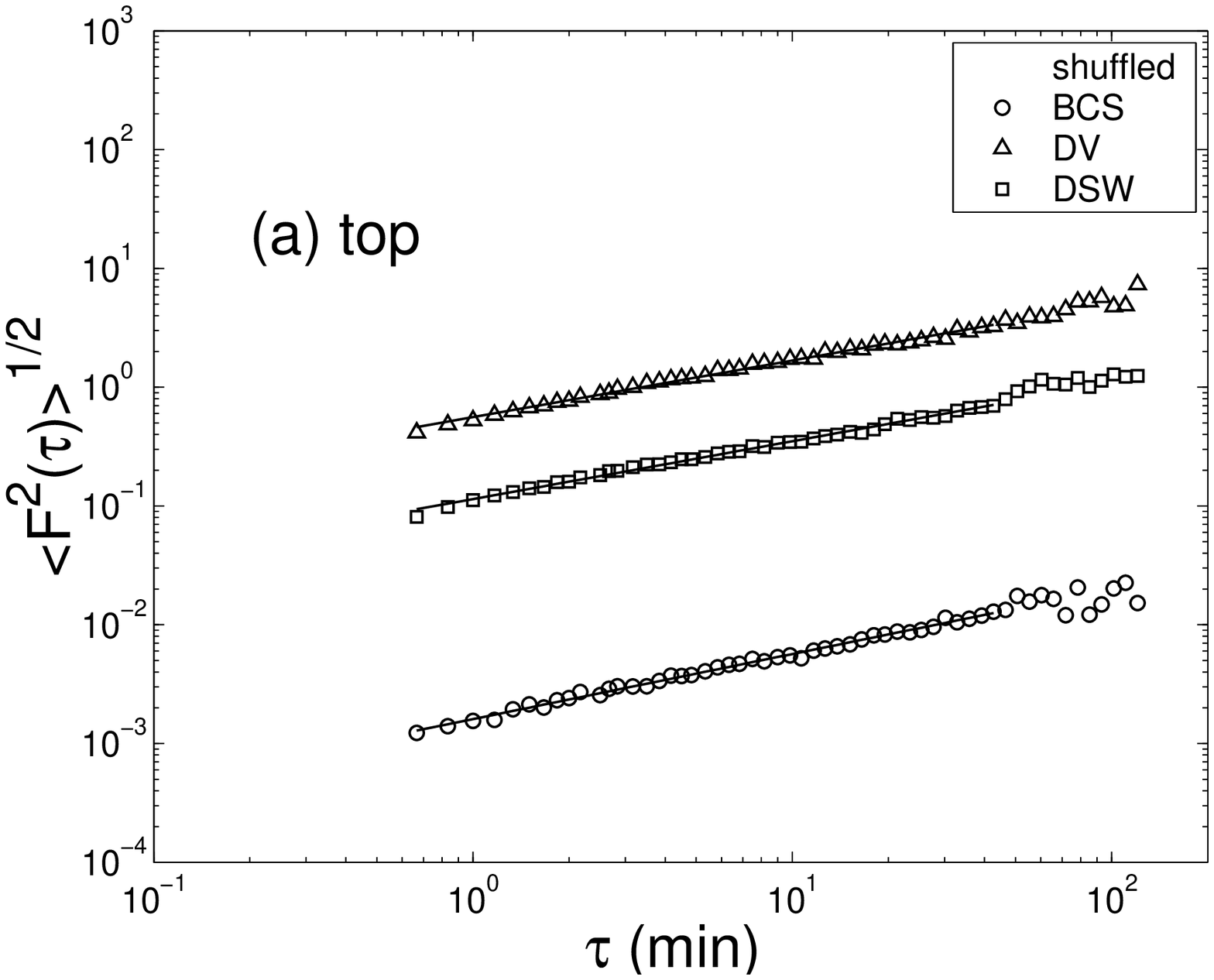} \hfill \leavevmode \epsfysize=4.5cm \epsffile{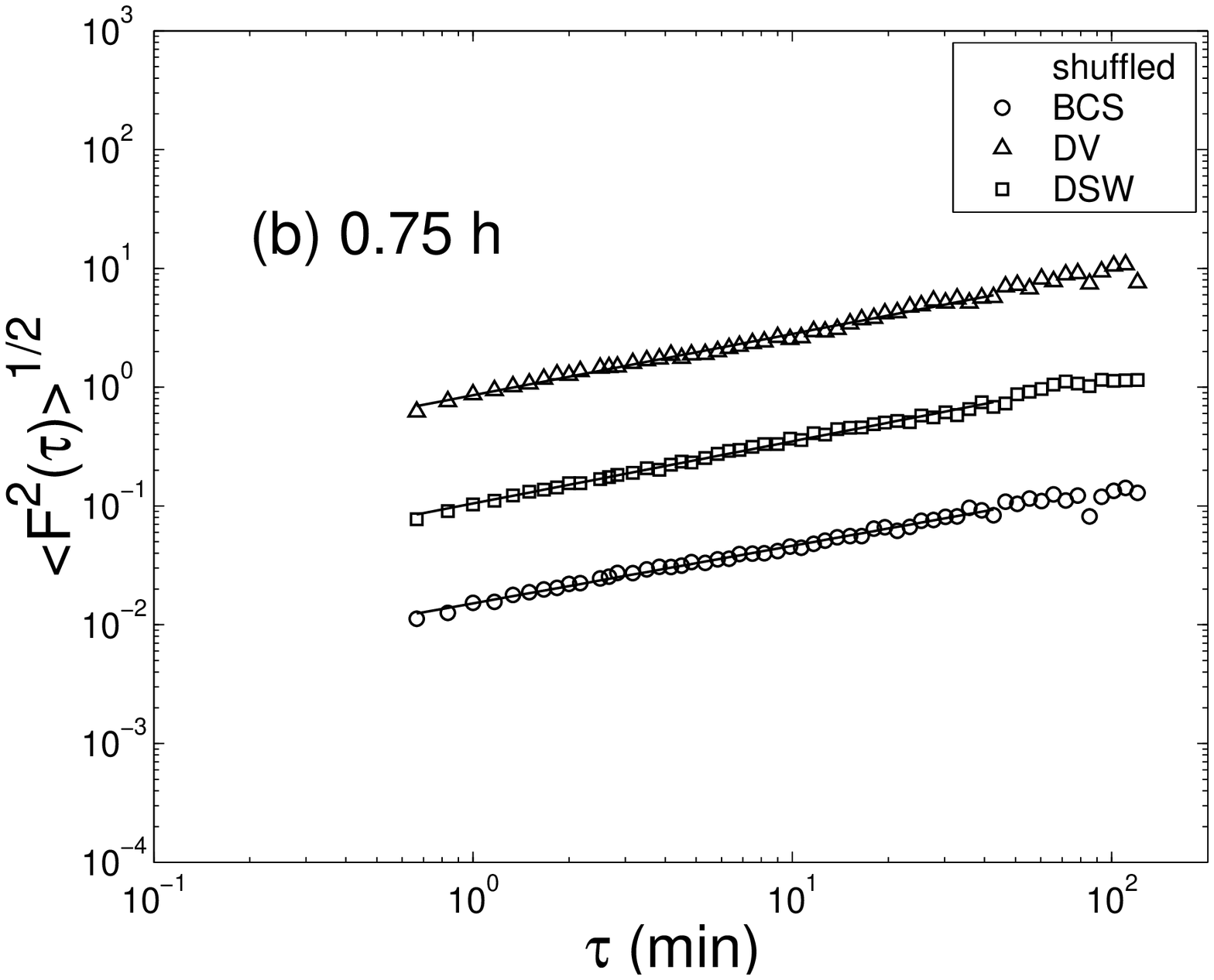}
\vfill \leavevmode \epsfysize=4.5cm \epsffile{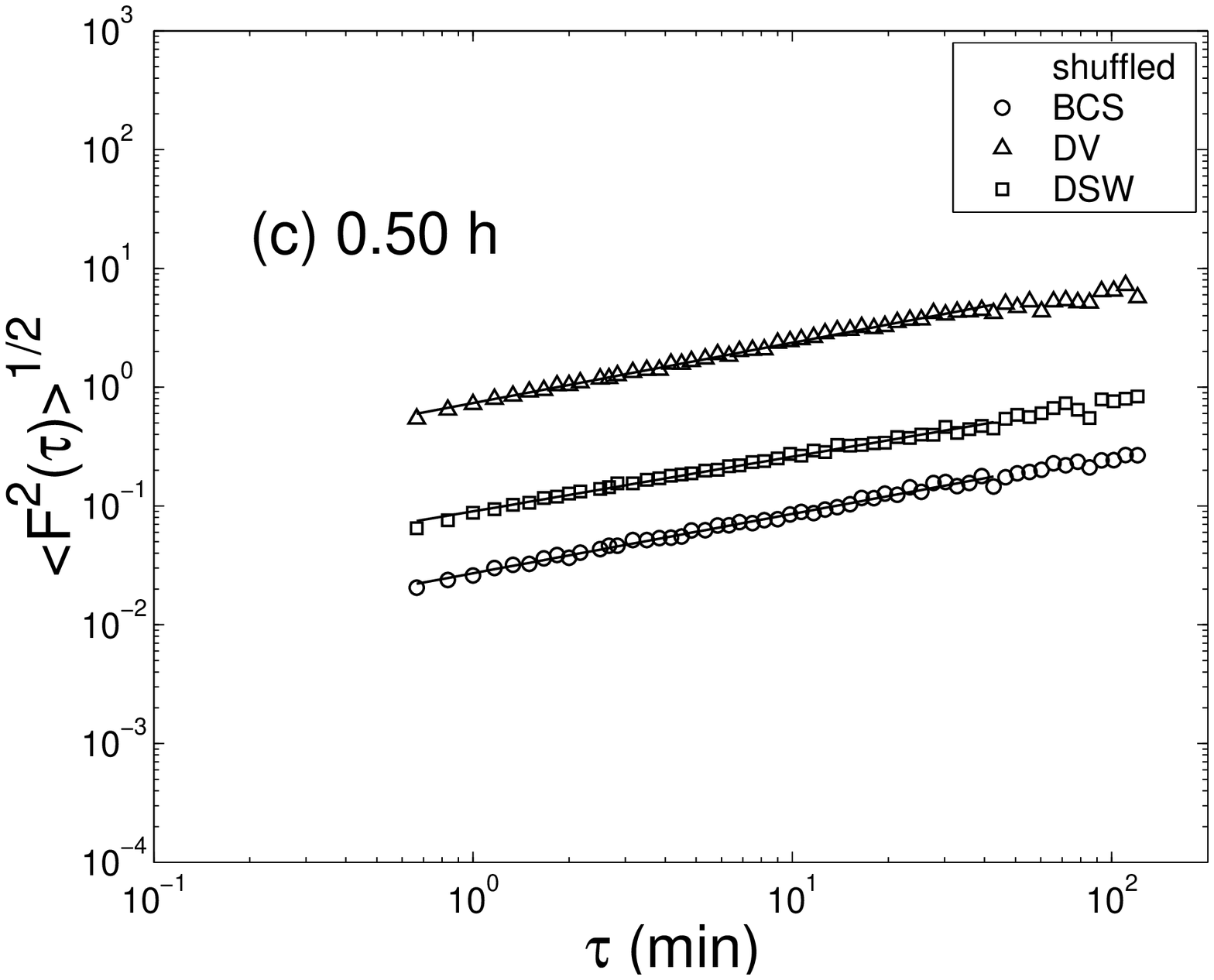} \vfill \leavevmode
\epsfysize=4.5cm \epsffile{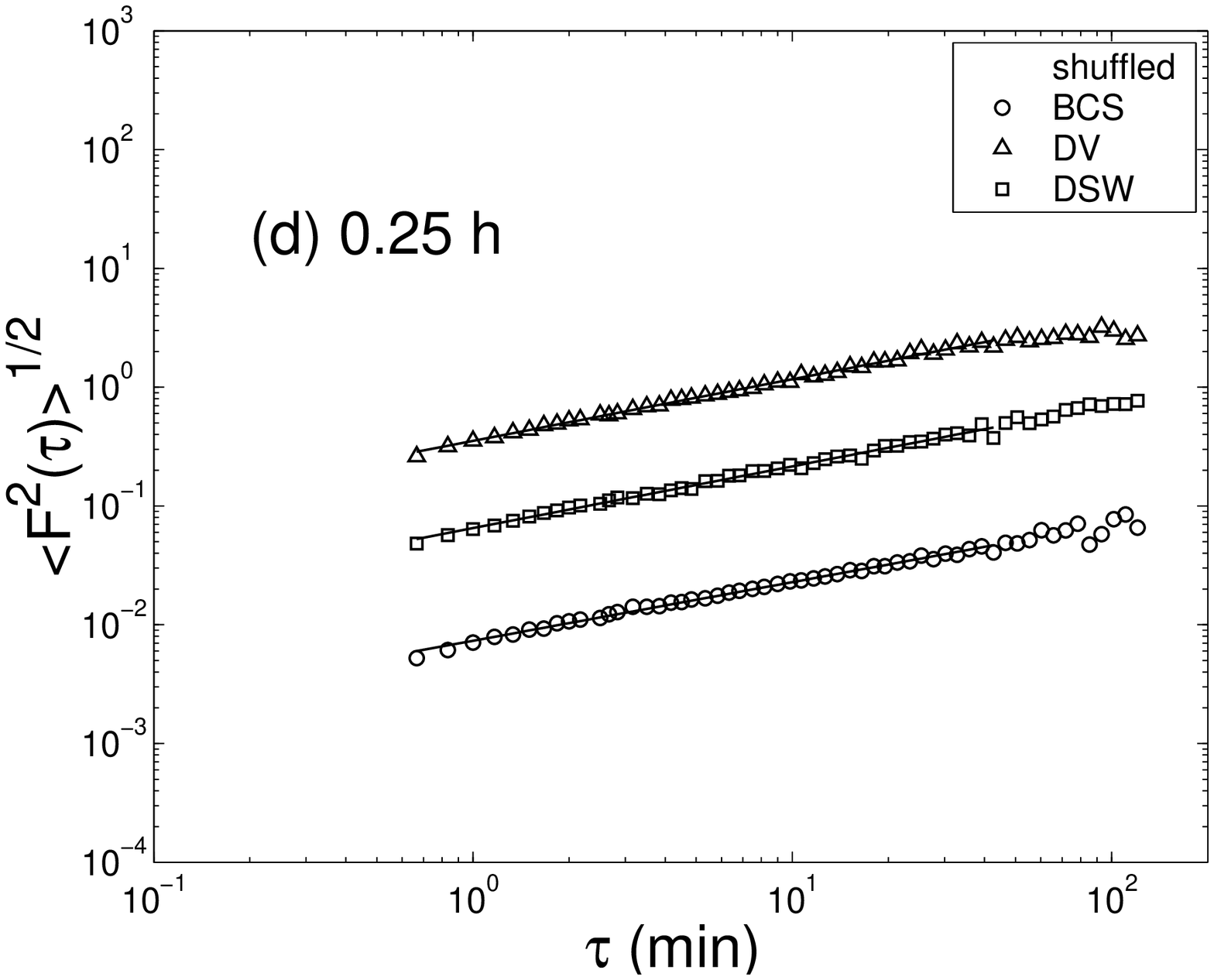} \hfill \leavevmode \epsfysize=4.5cm
\epsffile{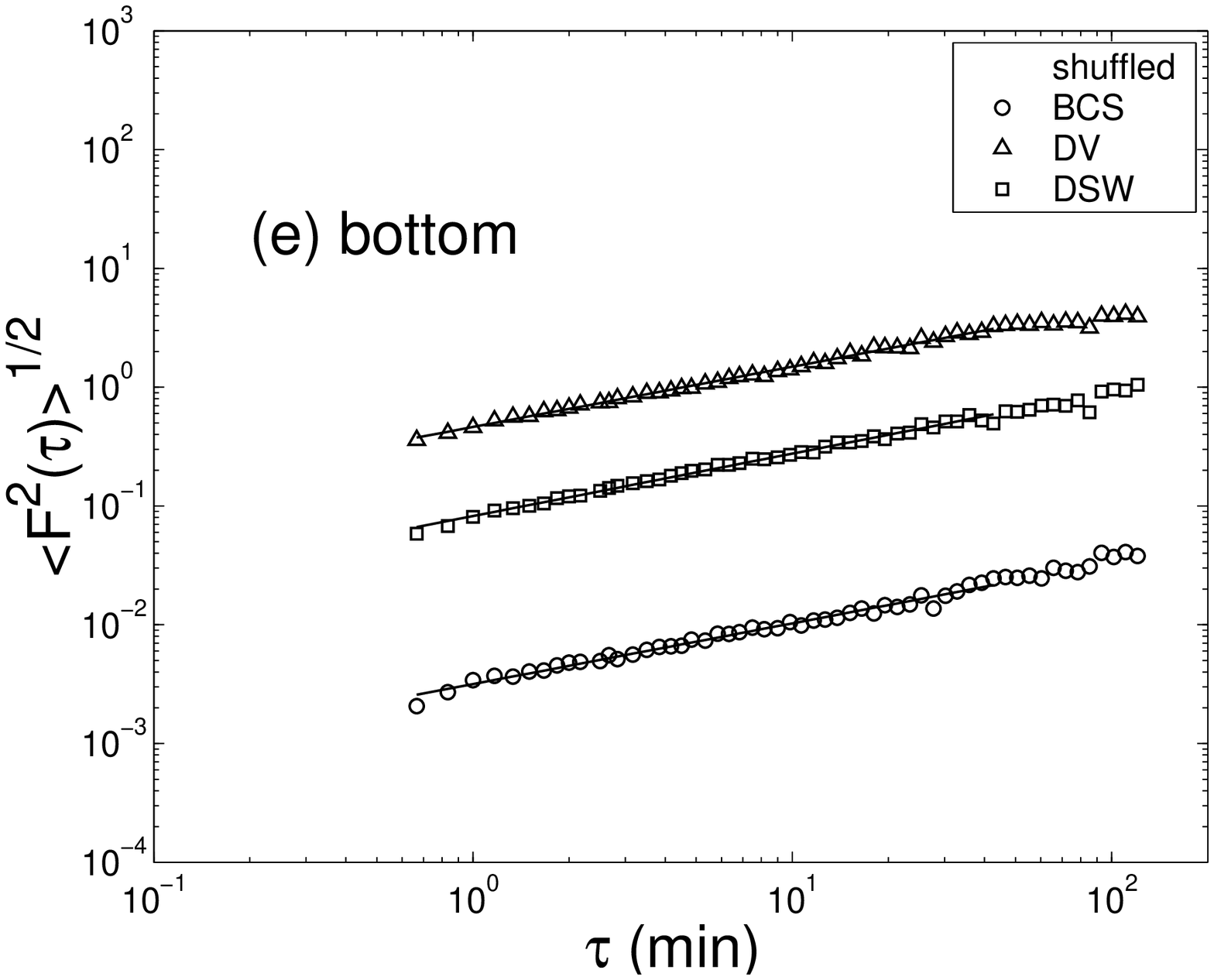} \end{center} \label{fig9} \caption{DFA-functions for the
shuffled data of: backscattering cross section (circles), Doppler velocity
(triangles) and Doppler spectral width (squares) at the (a) top, (b) 0.75, (c)
0.50, (d) 0.25 relative to the thickness $h$ of the cloud, and (e) 
bottom of the
$fall$ cloud shown in Fig. 3. DFA-functions are displaced for readability. The
$\alpha$-values are listed in Table 4.} \end{figure}

\end{document}